# Resuspension of Small Particles from Multilayer Deposits in Turbulent Boundary Layers

F. Zhang[†,*,♠], M. Reeks[*], M. Kissane[†] and R. J. Perkins[♠]

[†] Institut de Radioprotection et de Sûreté Nucléaire, BP 3, 13115 St-Paul-lez-Durance, France

[*] School of Mechanical and Systems Engineering, Newcastle University, Newcastle upon Tyne, NE1 7RU, UK

[♠] Laboratoire de Mécanique des Fluides et d'Acoustique, Ecole Centrale de Lyon, 36, avenue Guy de Collongue 69134 Ecully, France

## Abstract

We present a hybrid stochastic model for the resuspension of micron-size particles from multilayer deposits in a fully-developed turbulent boundary layer. The rate of removal of particles from any given layer depends upon the rate of removal of particles from the layer above which acts as a source of uncovering and exposure of particles to the resuspending flow. The primary resuspension rate constant for an individual particle within a layer is based on the Rock'n'Roll (R'n'R) model using non-Gaussian statistics for the aerodynamic forces acting on the particles (Zhang et al., 2012). The coupled layer equations that describe multilayer resuspension of all the particles in each layer are based on the generic lattice model of Friess & Yadigaroglu (2001) which is extended here to include the influence of layer coverage and particle size distribution. We consider the influence of layer thickness on the resuspension along with the spread of adhesion within layers, and the statistics of non-Gaussian versus Gaussian removal forces including their timescale. Unlike its weak influence on long-term resuspension rates for monolayers, this timescale plays a crucial and influential role in multilayer resuspension. Finally we compare model predictions with those of a large-scale and a mesoscale resuspension test, STORM (Castelo *et al.*, 1999) and BISE (Alloul-Marmor, 2002).

## 1. Introduction

In a previous paper (Zhang *et al.*, 2012) we presented an improved version of the Rock'n'Roll model (Reeks & Hall, 2001) for the resuspension of deposited particles by a turbulent flow. This model employs a stochastic approach to resuspension involving the rocking and rolling of a particle about surface asperities induced by the moments of the fluctuating drag force acting on the particle close to the surface. It represented a significant improvement over the original model because it used a joint distribution of the moments of the drag force $f(t)$ and its derivative $\dot{f}(t)$ acting on the particle based on DNS data of the near-wall stream-wise fluid velocity and acceleration in turbulent channel flow. Unlike the original model in which both $f(t)$ and $\dot{f}(t)$ are assumed to have Gaussian statistics, this improved non-Gaussian model accounted for the influence of the highly non-Gaussian forces (associated with the sweeping and ejection events in a turbulent boundary layer) on the resuspension rate.

Analysis and model predictions were focused on the resuspension of small particles from microscopically rough surfaces where even for a nominal macroscopically-smooth surface there is a broad range of surface molecular adhesive forces. In all cases the coverage of the surface with particles was assumed to be less than a monolayer, so that individual particles could be considered as isolated, i.e., attached only to the surface substrate and not to other particles. An important feature of resuspension from rough surfaces is that the resuspension takes place in two distinguishable phases, an initial short-term resuspension and a longer-term resuspension akin to erosion. The short-term resuspension takes places over a period ~ timescale of the fluctuating removal forces when typically the aerodynamic force ~ mean adhesive force and a significant fraction of the particles on the surface may be detached. Those particles that remain on the surface after this initial phase are much more strongly attached to the surface and are resuspended on a much longer timescale. In fact the resuspension (erosion) rate for these particles is found to vary almost inversely as the exposure time and is only very weakly dependent upon the timescale of the removal forces. Furthermore whilst the absolute decay rate is dependent upon the spread and mean of the adhesive forces/aerodynamic forces, these have little effect on the inverse time decay. These are features that are characteristic of stochastic approach and the existence of a a broad range of adhesive forces and we investigated the dependence of the long-term erosion rate on the spread of the adhesion and the distribution of the removal forces about their mean value (which was strongly non-Gaussian)

In this paper we focus on multilayer resuspension, i.e., the resuspension of particles from multilayer deposits of aerosol particles by a turbulent flow. This is the most general case for resuspension of deposited particles and the one most likely to arise in a severe nuclear accident. It is also a common feature of large-scale severe accident experiments involving both deposition and resuspension of aerosol particles, e.g., the STORM experiments (Castelo *et al.*, 1999). To be more precise, we use the formula for the resuspension rate constant given by the non-Gaussian R'n'R model in a hybrid model for the multilayer resuspension based on the approach used by Friess and Yadigaroglu (2001). We then examine how resuspension depends upon a number of key parameters: the number of layers, the distribution of adhesive forces and particle size, coverage and packing within each layer, the influence of non-Gaussian versus Gaussian removal forces and, perhaps most important of all, their timescale. Unlike their weak influence on long-term resuspension rates for monolayers, the timescales associated with the removal forces play a crucial and influential role in multilayer resuspension. Throughout this study we will compare our predictions with those of a monolayer. Most importantly we will also compare model predictions with those of two large-to-medium scale resuspension tests, namely STORM and BISE

We begin in the next section with a brief description of the non Gaussian R'n'R model focusing on the form given for the 'primary' resuspension rate constant. We then follow this in section 3 with a description of various multilayer models that have been developed and used in the past, the particular focus being on the Friess and Yadigaroglu (FY) multilayer generic model which is used as the basis of the hybrid model we present and study here. In this section we also show how the R'n'R model is incorporated into the FY model and present the set of equations for the resuspension of particles from each layer. In section 4 we consider their solution for the resuspension rate and fraction resuspended and its dependence on the various parameters we have referred to previously. In sections 5 and 6 we adapt the model to incorporate the influence of coverage (section 5) and a polydispersed deposit (section 6) i.e. a particle size distribution in each layer. Then in section 7 we show how these predictions compare with the actual experimental results for the fraction resuspended in the two multilayer resuspension experiments, STORM (Castelo *et al.*, 1999) and BISE (Alloul-Marmor, 2002). Section 8 concludes with a summary and conclusions.

## 2. The non-Gaussian Rock 'n'Roll (R'n'R) Model, a Brief Description

The geometry of the particle-surface contact in the R'n'R model is shown in Figure 1b in which the distribution of asperity contacts is reduced to a simple two-dimensional model of two-point asperity contact. Thus, rather than the centre of the particle oscillating vertically as in the original Reeks, Reed and Hall (1988) (RRH) model, it will oscillate/ rock about the pivot P until contact with the other asperity at Q is broken. When this happens it is assumed that the lift force is either sufficient to break the contact at P and the particle resuspends or it rolls until the adhesion at single-point contact is sufficiently low for the particle to resuspend. In either situation the rate of resuspension is controlled by the rate at which contacts are initially broken.

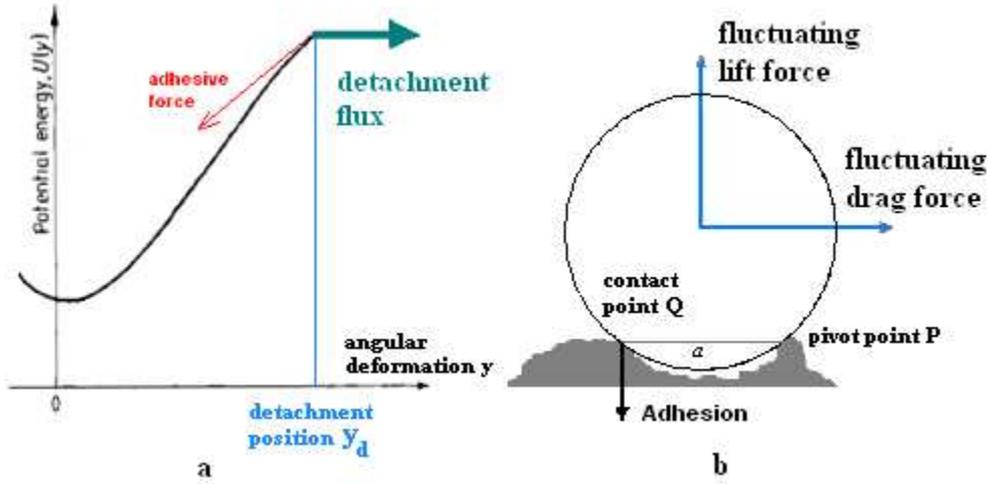

**Figure 1 -** Potential well and particle couple system

The formula for the resuspension rate has the same form as in the original RRH model except that couples are taken account of by replacing vertical lift forces by equivalent forces based on their moments. That is, referring to Figure 1b, the equivalent force $F$ is derived from the net couple ($\Gamma$) of the system above so that

$$\Gamma = \frac{a}{2}F_L + rF_D \quad \Rightarrow \quad F = \frac{1}{2}F_L + \frac{r}{a}F_D \qquad [1]$$

where $a$ is the typical distance between asperities, $r$ the particle radius, $F_L$ the lift force and $F_D$ the drag force. The ratio $r/a \sim 100$ (based on Hall's experiment, Reeks & Hall 2001) meaning that drag plays the dominant role in particle removal. We recall that for the quasi-static case in the R'n'R model, at the detachment point (i.e., point $y_d$ in Figure 1a, referring to the angular displacement of the asperity contact at Q about P as in Fig 1b), the aerodynamic force acting on the particle (which includes the mean $<F>$ and fluctuating parts $f(t)$) is considered to balance the restoring force at each instant of time (hence the term quasi-static). So,

$$\langle F \rangle + f(t) + F_A(y) = 0 \qquad [2]$$

where $F_A(y)$ is the adhesive restoring force as a function of the angular deformation ($y$) of the particle. At the point of detachment ($y_d$) the adhesive pull off f (the force required to detach the particle is = $-F_A(y_d)$. Following the tradition of previous authors we refer to this force as the force of adhesion. In the presence of applied mean force $\langle F \rangle$ from Eq.[2], the value of the fluctuating component of the equivalent aerodynamic force required to detach the particle ($f_d$) is given by

$$f_d = f_a - \langle F \rangle \qquad [3]$$

So as $F(t)$ fluctuates in time, the adhesive force $F_A(y)$ and hence $y(t)$ changes to balance it according to Eq.[2]; every time the value of $f(t)$ exceeds the value of $f_d$ a particle is detached from the surface. So the rate of detachment depends not only on the value of $f_d$ but on the frequency at which it is exceeded, i.e., upon the typical timescale of the fluctuating aerodynamic force $f(t)$ and its distribution in time.

Based on their measurements, the mean drag and lift force for a spherical particle of radius $r$ is given by Reeks and Hall (2001) as

$$\langle F_D \rangle = 32 \rho_f v_f^2 \left( \frac{r u_\tau}{v_f} \right)^2 \qquad \langle F_L \rangle = 20.9 \rho_f v_f^2 \left( \frac{r u_\tau}{v_f} \right)^{2.31} \qquad [4]$$

where $\rho_f$ is the fluid density, $v_f$ the fluid kinematic viscosity, and $u_\tau$ the wall friction velocity. The adhesive force is considered as a scaled reduction of the adhesive force on a smooth surface based on the JKR model (Johnson, Kendal and Roberts, 1971). Thus,

$$f_a = \frac{3}{2} \pi \gamma r r'_a \qquad [5]$$

where $\gamma$ is the surface energy and $r_a'$ the normalised asperity radius $r_a / r$ where $r_a$ is the asperity radius. $r_a'$ is assumed to have a log-normal distribution $\varphi(r_a')$ with geometric mean $\bar{r}'_a$ and geometric standard deviation $\sigma'_a$. Physically, these two parameters define the microscale roughness of the surface. $\bar{r}'_a$ is a measure of how much the adhesive force is reduced from its value for smooth contact with a surface and $\sigma'_a$ describes how broad/narrow the distribution is. For convenience we call $\bar{r}'_a$ the reduction in adhesion and $\sigma'_a$ the spread. Hall's experimental measurements of the distribution of adhesive forces on a polished stainless steel surface gave values of $\bar{r}'_a$ ~0.01 and a spread $\sigma'_a$ ~3. For a log-normal distribution $\varphi(r_a')$ is given explicitly by

$$\varphi(r'_a) = \frac{1}{\sqrt{2\pi}} \frac{1}{r'_a} \frac{1}{\ln \sigma'_a} \exp\left( -\frac{[\ln(r'_a / \bar{r}'_a)]^2}{2(\ln \sigma'_a)^2} \right) \qquad [6]$$

Biasi *et al.* (2001) took the R'n'R model for resuspension and added an empirical log-normal distribution of adhesive forces to reproduce the resuspension measurements of a number of experiments. Some adhesion-force parameters were tuned to fit the data of the most highly-characterised experiments, i.e., those of Hall (Reeks & Hall, 2001) and Braaten (1994). Then, for an enlarged dataset including STORM and ORNL ART resuspension results, the best global correlations for geometric mean adhesive force and geometric spread as a function of particle geometric mean radius (in microns) were obtained, namely

$$\bar{r}'_a = 0.016 - 0.0023 r^{0.545}$$
$$\sigma'_a = 1.8 + 0.136 r^{1.4} \qquad [7]$$

where $r$ is the particle radius in microns.

The resuspension rate constant $p$, according to Reeks *et al.* (1988), is defined as the number of particles per second detached from the surface over the number of particles attached to the surface.

$$p = \int_0^\infty v P(y_d, v) dv \bigg/ \int_{-\infty}^\infty \int_{-\infty}^{y_d} P(y, v) dy dv \qquad [8]$$

where $y$ is the displacement or deformation of the centre of the particle, $v = dy/dt = \dot{y}$ and $P$ is the joint distribution of $v$ and $y$. The numerator is the particle detachment flux at the point of detachment (Figure 1a) whilst the denominator is the number of particles attached to the surface, i.e., in the potential well.

Referring to Eq.[2] for the quasi-static case, we note that the angular deformation or displacement $y$ can be written as an implicit function of the fluctuating aerodynamic force, $f(t)$, i.e.,

$$y(t) = \psi(f) \text{ and so } \dot{y}(t) = \dot{f}\psi'(f) \quad [9]$$

where $\psi'(f)$ is the first derivative of $\psi(f)$ with respect to $f$.
Then

$$p = \int_0^\infty \dot{f} P(f_d, \dot{f}) d\dot{v} \bigg/ \int_{-\infty}^\infty \int_{-\infty}^{f_d} P(f, \dot{f}) df d\dot{f} \,] \quad [10]$$

where in the R'n'R model the joint distribution $P$ of fluctuating aerodynamic force $f$ and its derivative $\dot{f}$ is assumed to be a joint normal distribution with zero correlation between the force and its derivative. In the modified R'n'R model (Zhang, 2011), a new joint distribution is developed from DNS data which compounded of a Rayleigh distribution for $z_1$ and a Johnson SU distribution for $z_2$. More precisely,

$$P(z_1, z_2) = \frac{z_1 + A_1}{A_2^2} \exp\left(-\frac{1}{2}\left(\frac{z_1 + A_1}{A_2}\right)^2\right) \cdot \frac{B_1}{B_2\sqrt{2\pi}\sqrt{z^2+1}} \exp\left(-\frac{1}{2}\left(B_3 + B_1 \ln\left(z + \sqrt{z^2+1}\right)\right)^2\right) \quad [11]$$

where $A_1, A_2, B_1, B_2, B_3$ and $B_4$ are constants depending on the wall distance $y^+$

$z_1$ and $z_2$ is the fluctuating aerodynamic force and derivative normalized on their r.m.s values, so

$$z_1 = \frac{f}{\sqrt{\langle f^2 \rangle}}, \quad z_2 = \frac{\dot{f}}{\sqrt{\langle \dot{f}^2 \rangle}} \quad [12]$$

Then the modified resuspension rate constant is obtained

$$p = \sqrt{\frac{\langle \dot{f}^2 \rangle}{\langle f^2 \rangle}} \int_0^\infty z_2 P(z_d, z_2) dz_2 \bigg/ \int_{-\infty}^\infty \int_{-\infty}^{z_d} P(z_1, z_2) dz_1 dz_2$$

$$= B_{\dot{f}} \sqrt{\frac{\langle \dot{f}^2 \rangle}{\langle f^2 \rangle}} \frac{z_d + A_1}{A_2^2} \exp\left(-\frac{1}{2}\left(\frac{z_d + A_1}{A_2}\right)^2\right) \bigg/ \left[1 - \exp\left(-\frac{1}{2}\left(\frac{z_d + A_1}{A_2}\right)^2\right)\right] \quad [13]$$

$$z_d = \frac{f_d}{\sqrt{\langle f^2 \rangle}}.$$

In the original R'n'R model the term $\sqrt{\langle f^2 \rangle}$ is expressed as a fraction $f_{rms}$ of the mean aerodynamic force, i.e.,

$$\sqrt{\langle f^2 \rangle} = f_{rms} \langle F \rangle \quad [14]$$

As for the original R'n'R model we write

$$\sqrt{\frac{\langle \dot{f}^2 \rangle}{\langle f^2 \rangle}} = \omega^+ \left(\frac{u_\tau^2}{\nu_f}\right) \quad [15]$$

Values for the various dimensionless parameters associated with the formula for $p$ in Eq.[13] are given below for values of $y^+ = 0.1$ for DNS measurements.

| DNS | $B_{\dot{f}}$ | $A_1$ | $A_2$ | $\omega^+$ | $f_{rms} = \sqrt{\langle f^2 \rangle}/\langle F \rangle$ |
|---|---|---|---|---|---|

| $y^+ = 0.1$ | 0.3437 | 1.8126 | 1.4638 | 0.1642 | 0.366 |

**Table 1** - Values of parameters used in formula for resuspension rate constant *p*

## 3. Multilayer Resuspension Models

Resuspension of multilayer deposits of radioactive particles is an important phenomenon in nuclear severe accidents. However, most of the models used to predict the amount of resuspended particulate are based on the resuspension of isolated particles. There are only a few models which consider the multilayer case, namely Fromentin (1989), Lazaridis & Drossinos (1998) (referred to as LD) and Friess & Yadigaroglu (2001) (referred to as FY). In this paper, the modified R'n'R model will be adapted for application to multilayer deposits based on the FY multilayer approach. Furthermore, the coverage effect of layers on resuspension will be considered in this modified FY multilayer model in two ways: 1) introducing a coverage factor; 2) considering the influence of a distribution of particle size within each layer. Finally, the modified multilayer model resuspension predictions will be compared with the resuspension measurements in the STORM SR11 test (Castelo *et al.*, 1999) and the BISE experiment (Alloul-Marmor, 2002).

Fromentin's model is a semi-empirical model based on force-balance methods and the input parameters are limited to those in the PARESS experiment upon which the model is based. The LD model is restricted to a multilayer deposit in which the resuspension rate for each particle in the deposit is the same. In contrast the FY model (2001) takes account of the fact that the resuspension rate constant can vary according to the distribution of adhesive forces experienced by a particle in each layer. If we compare the two models in situations where the resuspension rate constant is the same for all particles, there is a difference in the way the fraction of particles exposed to the flow is calculated. If the layers are numbered 1, 2, 3...etc from the top layer exposed to the flow, then the LD model takes this fraction in the *i*th layer to be the number of particles removed from the (*i*-1)th layer over the initial number of particles in the (*i*-1)th layer. Thus it assumes that whenever a particle is removed from the (*i*-1)th layer a particle in the *i*th layer is exposed and is immediately resuspended. Hence, the model is valid for large resuspension rates or, equivalently, it gives the maximum multilayer resuspension rate (i.e., if a particle is exposed it resuspends). In these circumstances the FY model, on the other hand, takes the fraction of exposed particles in the *i*th layer to be the ratio of the number of particles in the (*i*-1)th layer to the number of particles in the *i*th layer at time *t*. Thus

$$\frac{dn_i}{dt} = -pn_i \left[ 1 - \frac{n_{i-1}(t)}{n_{i-1}(0)} \right] \quad i \geq 2 \quad \text{LD model}$$

$$\frac{dn_i}{dt} = -pn_i \left[ 1 - \frac{n_{i-1}(t)}{n_i(t)} \right] \quad i \geq 2 \quad \text{FY model}$$

where $n_i$ is the number of particles in the *i*th layer at time *t* (i.e., the total number of particles in *i*th layer at time *t* being the sum of those particles exposed and unexposed to the flow) and *p* is the resuspension rate constant. Since every particle sits on top of a particle in the layer below, this ratio gives the exact number of exposed particles which the LD model does not provide.

The FY model is not only exact in this situation but, as we stated above, more generally applicable since the resuspension rate constant can vary from particle to particle in any given layer and from layer to layer. Since this is an important consideration in multilayer resuspension, the FY model is used as the basis for the multilayer modelling which will be presented here.

Let us now recall the essential features of the FY model (Friess & Yadigaroglu, 2001) which deals with the resuspension of a deposit composed of a regular array of identical spherical particles. FY (2001) first dealt with an infinitely thick deposit, defining a resuspension rate constant $p(\xi)$ for each particle exposed to the flow, where $\xi$ is a statistical variable or number of variables upon which $p$ depends (e.g., adhesion arising from the contact of a particle with other particles whose value controls the influence of the flow). The probability density function (pdf) for the occurrence of $\xi$ is given by $n(\xi,t)$. It is assumed that every time a particle is removed (resuspended by the flow) another particle is uncovered and exposed to the flow but with a different $\xi$ say $\xi'$ with a probability distribution $\varphi(\xi')$ for uncovered particles. The particles are homogeneously distributed in the initial state with $n(\xi,0) = \varphi(\xi)$. The rate at which particles are exposed is given by

$$\Lambda(t) = \int_\xi p(\xi)n(\xi,t)d\xi \qquad [16]$$

and the fraction of those particles exposed per unit time with values between $\xi, \xi + d\xi$ will be

$$d\xi\varphi(\xi)\int_{\xi'} p(\xi')n(\xi',t)d\xi'$$

The equation for $n(\xi,t)$ is thus

$$\frac{\partial n(\xi,t)}{\partial t} = -p(\xi)n(\xi,t) + \varphi(\xi)\int_{\xi'} p(\xi')n(\xi',t)d\xi' \qquad [17]$$

Therefore, for an L-layer deposit, let $n_i(\xi,t)d\xi$ denote the probability of exposed particles between $\xi, \xi + d\xi$ in the $i$th layer at time $t$, the layers being numbered sequentially from the top layer (totally exposed to a flow) downward as $i = 1, 2, 3 \ldots$ L. Then the set of ODEs (ordinary differential equations) is

$$\frac{\partial n_1(\xi,t)}{\partial t} = -p(\xi)n_1(\xi,t)$$

$$\frac{\partial n_i(\xi,t)}{\partial t} = -p(\xi)n_i(\xi,t) + \varphi(\xi)\int_{\xi'} p(\xi')n_{i-1}(\xi',t)d\xi' \quad (i \geq 2) \qquad [18]$$

The resuspension rate for $i$th layer is given by

$$\Lambda_i(t) = \int_\xi p(\xi)n_i(\xi,t)d\xi \qquad [19]$$

**3.1 R'n'R Hybrid Model for Multilayer Resuspension**

*3.1.1 Formulation using the Friess & Yadigaroglu (FY) Generic Model (Hybrid Generic Model)*

The non-Gaussian R'n'R model is a single particle kinetic model based on the R'n'R model (Reeks & Hall, 2001) (with the resuspension rate constant $p(\xi_i)$ accounting for the particle-surface adhesive forces and the non-Gaussian statistics of the fluctuating aerodynamic resultant force obtained from DNS data).

We recall that in the FY multilayer model, this distribution of normalized asperity radii refers to the initial state of the pdf of exposed particles, $n(\xi,0)$, i.e.,

$$n(r_a',0) = \varphi(r_a') \qquad [22]$$

where the normalized asperity radius $r_a'$ refers to the intrinsic statistical variable $\xi$.

The resuspension rate constant $p$ in the modified R'n'R model (Eq.[13]) is a function of the normalized asperity radius for a fixed particle size. Hence, $p$ is referred to as $p(r_a')$. Therefore, the set of ODEs for the pdf of exposed particles in a deposit composed of $i = 1, 2, 3 \ldots L$ layers at time $t$ is given by

$$\frac{\partial n_1(r_a',t)}{\partial t} = -p(r_a')n_1(r_a',t)$$

$$\frac{\partial n_i(r'_a,t)}{\partial t} = -p(r'_a)n_i(r'_a,t) + \varphi(r'_a)\int_0^\infty p(\tilde{r}'_a)n_{i-1}(\tilde{r}'_a,t)d\tilde{r}'_a \quad (i \geq 2) \tag{23}$$

The $\varphi(r'_a)\int_0^\infty p(\tilde{r}'_a)n_{i-1}(\tilde{r}'_a,t)d\tilde{r}'_a$ part is thus a source term. It is considered as the extra source of particle exposure rate (exposing particles in a given layer by resuspending particles from the layer above).

The resuspension rate in the $i$th layer is given by

$$\Lambda_i(t) = \int_0^\infty p(r'_a)n_i(r'_a,t)dr'_a \tag{24}$$

For $i = 1$, the first layer resuspension rate is given explicitly in the original R'n'R model appropriate for a monolayer because there is no source term from the particles in the layer above, namely

$$\Lambda_1(t) = \int_0^\infty p(r'_a)e^{-p(r'_a)t}\varphi(r'_a)dr'_a \tag{25}$$

The initial value of exposed particles from the second layers downward $n_i(r_a',0) = 0$ ($i \geq 2$) which means all the particles in the layer below are covered by the particles from the layer above. Therefore, for the second layer and layers below ($i \geq 2$) using the initial condition for $n_i(r_a',t)$, gives

$$\left[\frac{\partial n_i(r'_a,t)}{\partial t} + p(r'_a)n_i(r'_a,t)\right] = \varphi(r'_a)\int_0^\infty p(\tilde{r}'_a)n_{i-1}(\tilde{r}'_a,t)d\tilde{r}'_a \quad (i \geq 2)$$

$$\Rightarrow e^{p(r'_a)t}\left[\frac{\partial n_i(r'_a,t)}{\partial t} + p(r'_a)n_i(r'_a,t)\right] = e^{p(r'_a)t}\varphi(r'_a)\int_0^\infty p(\tilde{r}'_a)n_{i-1}(\tilde{r}'_a,t)d\tilde{r}'_a$$

$$\Rightarrow e^{p(r'_a)t}n_i(r'_a,t) = \int_0^t e^{p(r'_a)t'}\varphi(r'_a)\int_0^\infty p(\tilde{r}'_a)n_{i-1}(\tilde{r}'_a,t')d\tilde{r}'_a dt'$$

$$\Rightarrow n_i(r'_a,t) = \varphi(r'_a)e^{-p(r'_a)t}\int_0^t e^{p(r'_a)t'}\int_0^\infty p(\tilde{r}'_a)n_{i-1}(\tilde{r}'_a,t')d\tilde{r}'_a dt' \quad (i \geq 2) \tag{26}$$

Substituting Eq.[26] in Eq.[24], the resuspension rate for the $i$th layer is given by

$$\Lambda_i(t) = \int_0^\infty p(r'_a)\varphi(r'_a)e^{-p(r'_a)t}\left[\int_0^t e^{p(r'_a)t'}\int_0^\infty p(\tilde{r}'_a)n_{i-1}(\tilde{r}'_a,t')d\tilde{r}'_a dt'\right]dr'_a \quad (i \geq 2) \tag{27}$$

Initially it is assumed as in the FY generic model that the deposit is formed from identical spherical particles as shown in Figure 2.

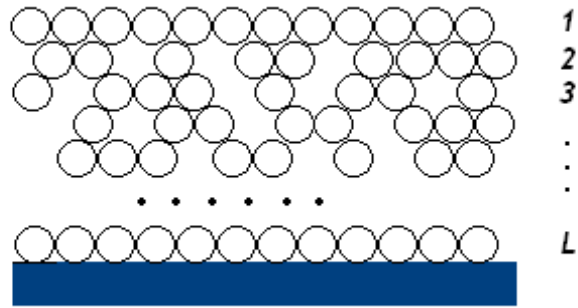

**Figure 2** - Geometry of the multilayer system with same-size particles

## 4. Hybrid Generic Model Predictions for Multilayer Resuspension

In this section we present the prediction of the hybrid generic multilayer model for the resuspension fraction and resuspension rates. In so doing the micro roughness for all the particles is assumed to be the same so the same distribution of normalized asperity radius $\varphi(r_a')$ is used for all the particles in the domain. In the calculation below, the parameters from the STORM SR11 experiment Phase 6 condition will be applied (more information is provided in the next section). The parameters used for Phase 6 are:

| Average radius ($\mu m$) | Fluid density ($kg.m^{-3}$) | Fluid kinematic viscosity ($m^2.s^{-1}$) | Wall friction velocity ($m.s^{-1}$) | Surface energy ($J.m^{-2}$) | Reduction factor (Geometric mean) | Spread factor (Geometric standard deviation) |
|---|---|---|---|---|---|---|
| 0.227 | 0.5730 | 5.2653 x $10^{-5}$ | 6.249 | 0.5 | 0.015 | 1.817 |

**Table 2** – Values of parameters in STORM SR11 Phase 6

where Biasi's correlation (Eq.[7]) is used for the adhesion distribution parameters. The particles in each layer are assumed to be monodisperse with a size given by the mean particle radius in Table 2.

### 4.1 Resuspension as a Function of Layer Number Location and Layer Thickness

Figure 3 shows the resuspension rate of particles in layers 1, 3, 5, 10 and 20. The resuspension rate of the first layer is identical to that of the isolated particle model. For the second and subsequent layers, the initial resuspension rate at $t = 0$ is zero. Starting from zero the resuspension rate rises to a maximum during which time most of the particles which are easily removed resuspend from a given layer (i.e., for these particles the mean effective aerodynamic force > adhesive force), the remainder being removed over a much longer period. The time to reach a maximum may therefore be regarded as a delay time for the particles in a given layer to be exposed by removing particles from all the layers above it. Note that the maximum resuspension rate decreases as the layer number increases. Also note that the resuspension rate of each layer beyond its maximum value, whilst being initially less than the layer above, eventually exceeds it so that the integrated amount (fraction resuspended) is the same (close to unity) for all layers in the long-term ($t \rightarrow \infty$) (the numerics were checked to make sure that this was the case).

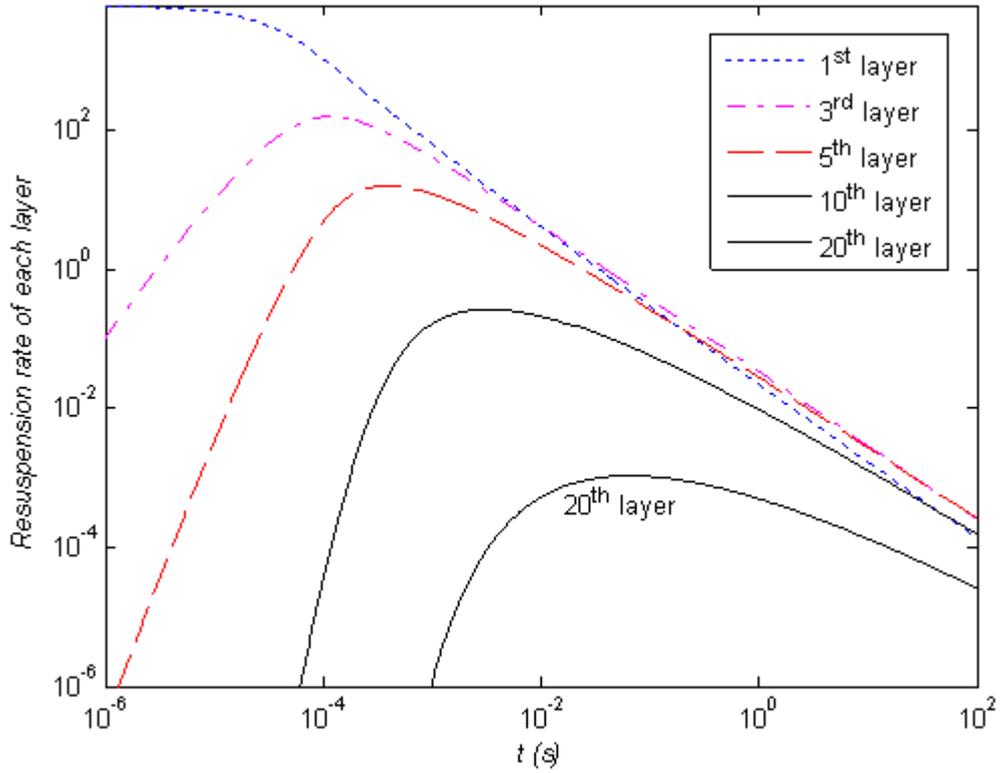

**Figure 3 -** Resuspension rate of each layer vs. time for the hybrid generic model using STORM test (SR11) Phase 6 conditions (see Table 2).

The initial number of particles is the same in each layer normalized to unity. Then the fractional resuspension rate for the domain (*L* layers) is given by

$$\Lambda_L(t) = \frac{\sum_{i=1}^{L} \Lambda_i(t)}{L} \qquad [28]$$

which corresponds to the resuspension fraction of an *L*-layer deposit.

$$f_r(t)_L = \int_0^t \Lambda_L(t') dt' \qquad [29]$$

The modified R'n'R model incorporating the FY multilayer approach will be referred to as the initial multilayer model since the model will later on be extended to include the coverage effect.

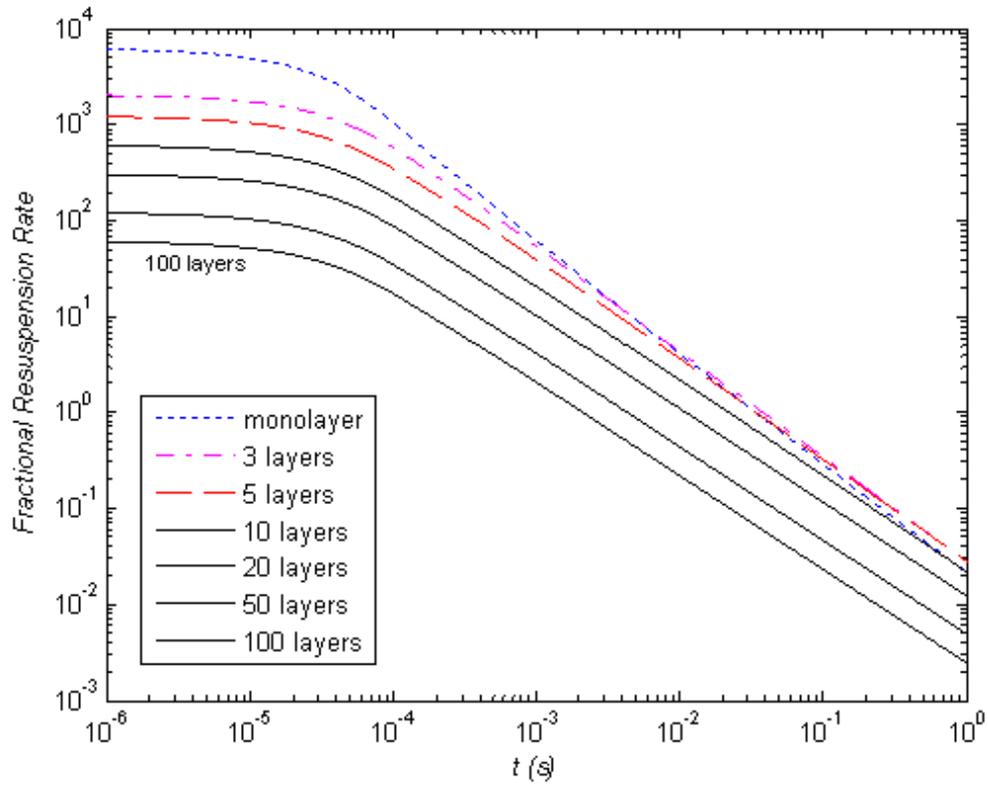

**Figure 4 -** Fractional resuspension rate as a function of number of layers (particle diameter: 0.45μm) based on STORM test (SR11) Phase 6 conditions (see Table 2).

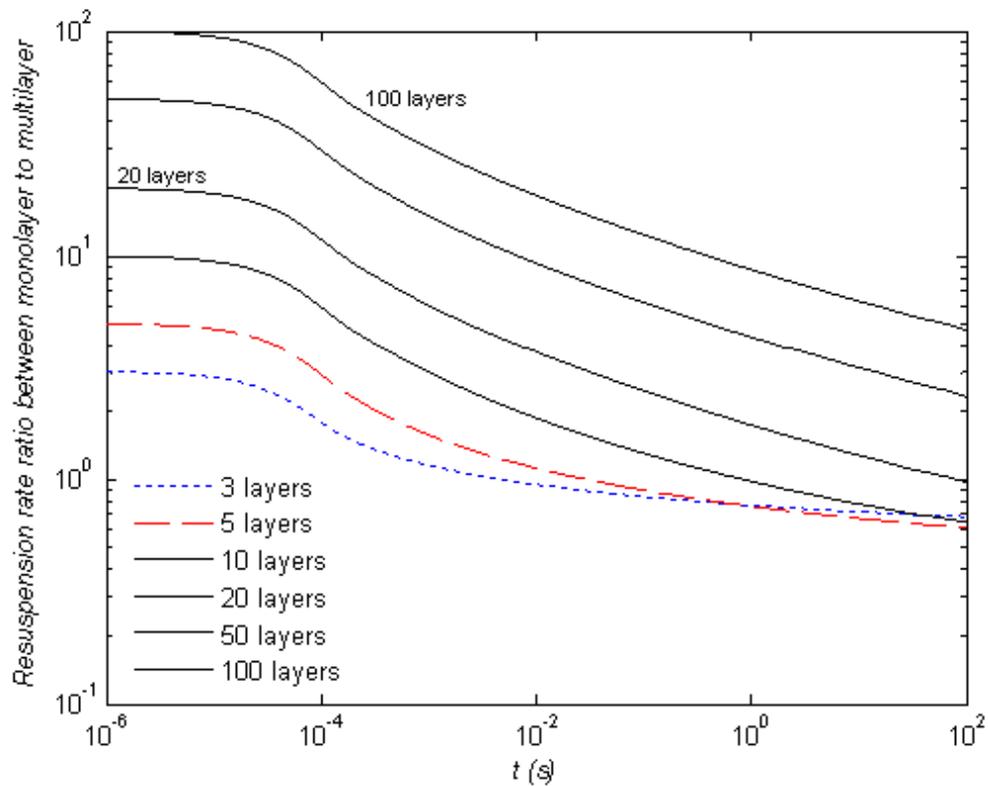

**Figure 5 -** Predictions of the hybrid generic model for ratio of monolayer / multilayer resuspension rates, based on STORM test (SR11) Phase 6 conditions (see Table 2).

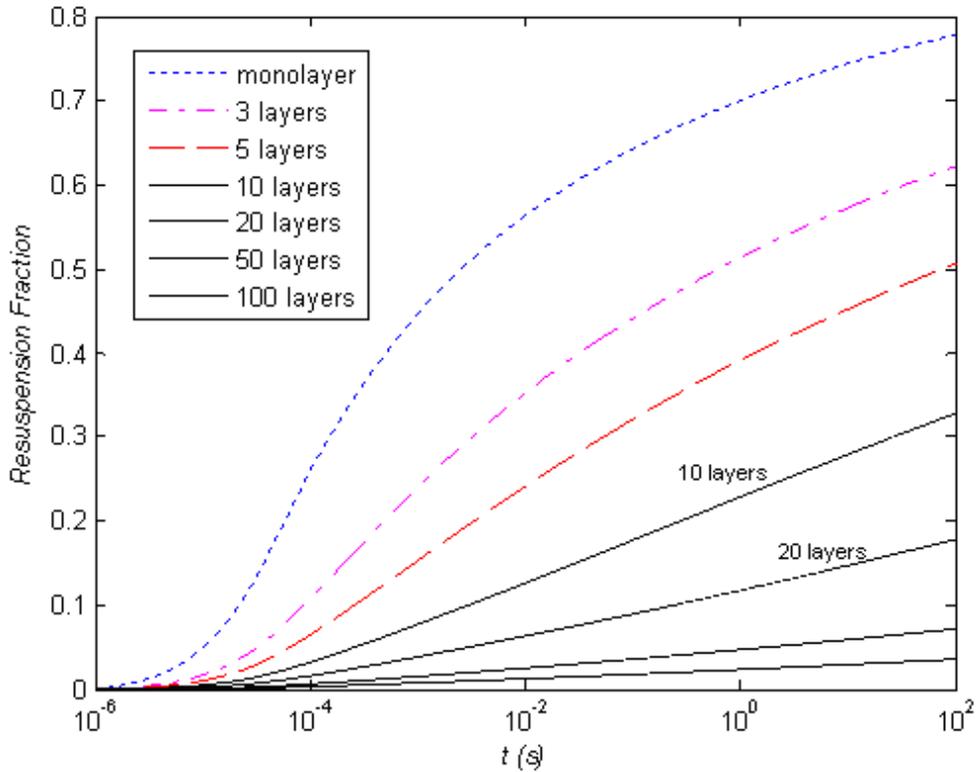

**Figure 6** - Hybrid generic model predictions for the resuspension fraction as a function of time and layer thickness based on STORM test (SR11) Phase 6 conditions (see Table 2).

The fractional resuspension rate is shown in Figure 4. The short term resuspension rate is considered finished when the exposure time $t$ is around $10^{-4}$s when as shown in the plot the long-term resuspension (approximately proportional to $1/t$) begins (we note that $1/\omega^+$ which corresponds to a timescale $\sim 10^{-5}$s). The initial fractional resuspension rate decreases with increasing layers when more particles tend to be removed over a longer period.

Figure 5 shows how much the monolayer fractional resuspension rate differs from that of the multilayer resuspension (for the same initial fraction = 1, in each case). As the layer number increases, the ratio of the resuspension rate between monolayer and multilayer increases significantly in the short term. However after a given time which increases with the value of $i$, the resuspension rate for a given layer eventually exceeds that for a monolayer $i = 1$ at the same time (the ratio < 1), and in the long-term the ratio should converge to zero (see the cases for layer $i = 3$ and 5 in Figure 3). Figure 6 shows the resuspension fraction of the whole deposit domain. It shows that after 100s 80% of the monolayer deposit is removed whereas only around 3% of the 100-layer deposit is resuspended.

**4.2 Influence of Adhesive Spread Factor on Multilayer Resuspension**

In these predictions we fix the geometric mean of the normalized asperity radius at 0.015. The spread factor values of 1.1, 1.817 (based on the Biasi correlation) and 4.0 are chosen for comparison (note that the spread is the geometric standard deviation and > 1). Figure 7, Figure 8 and Figure 9 show the comparison of resuspension fraction for a monolayer, 10-layer and 100-layer deposits for different spread factors, respectively. We note that for a monolayer, by virtue of the log-normal distribution, the time for 50% of the deposit to resuspend is independent of the spread factor. For times smaller than this value, the largest spread factor gives the greatest fraction resuspended whereas beyond this time, the situation is reversed.

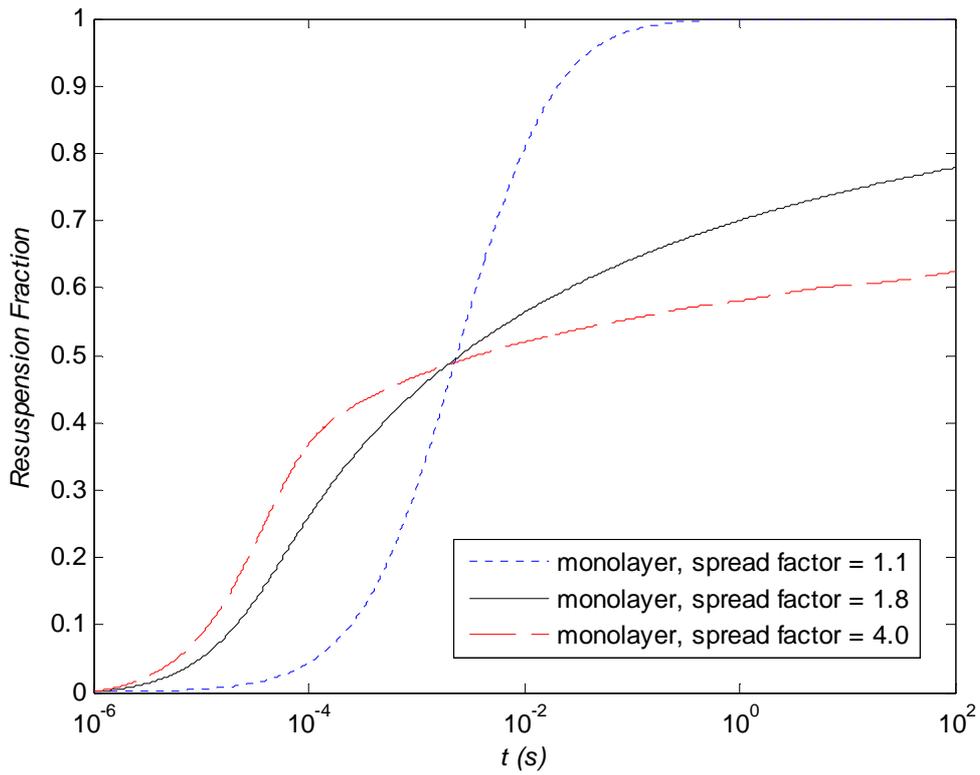

**Figure 7 -** Effect of adhesion spread factor on monolayer resuspension fraction (note that the exposure for 50% resuspension is the same for all spread factors (based on STORM test (SR11) Phase 6 conditions (see Table 2))

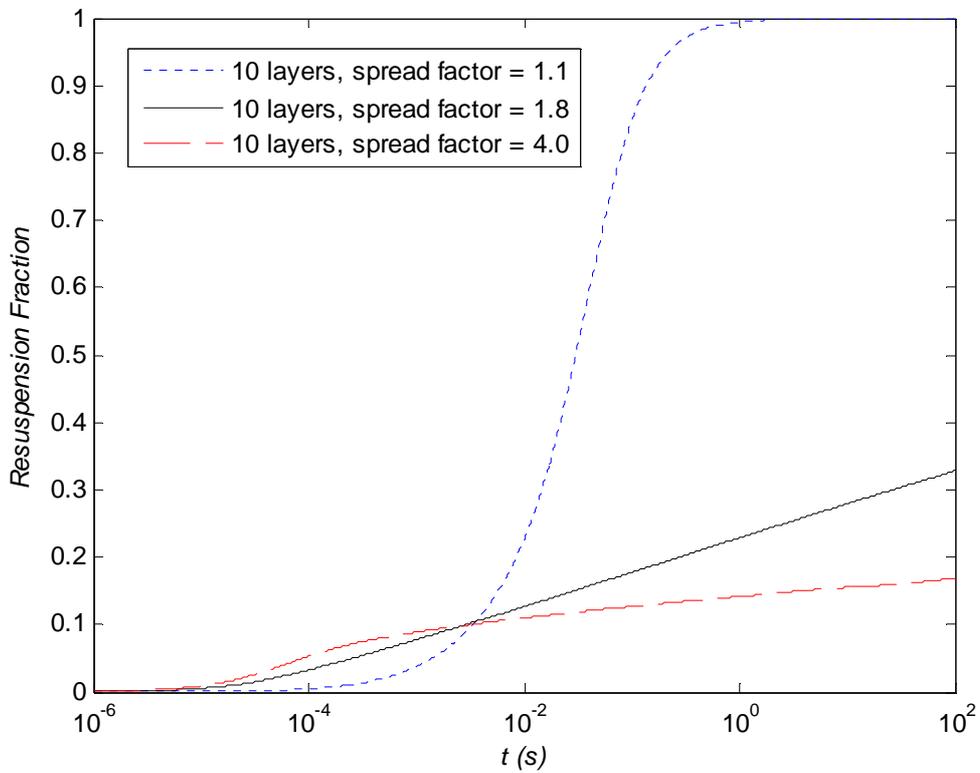

**Figure 8 -** Effect of adhesion spread factor on 10 layer deposit based on STORM test (SR11) Phase 6 conditions (see Table 2).

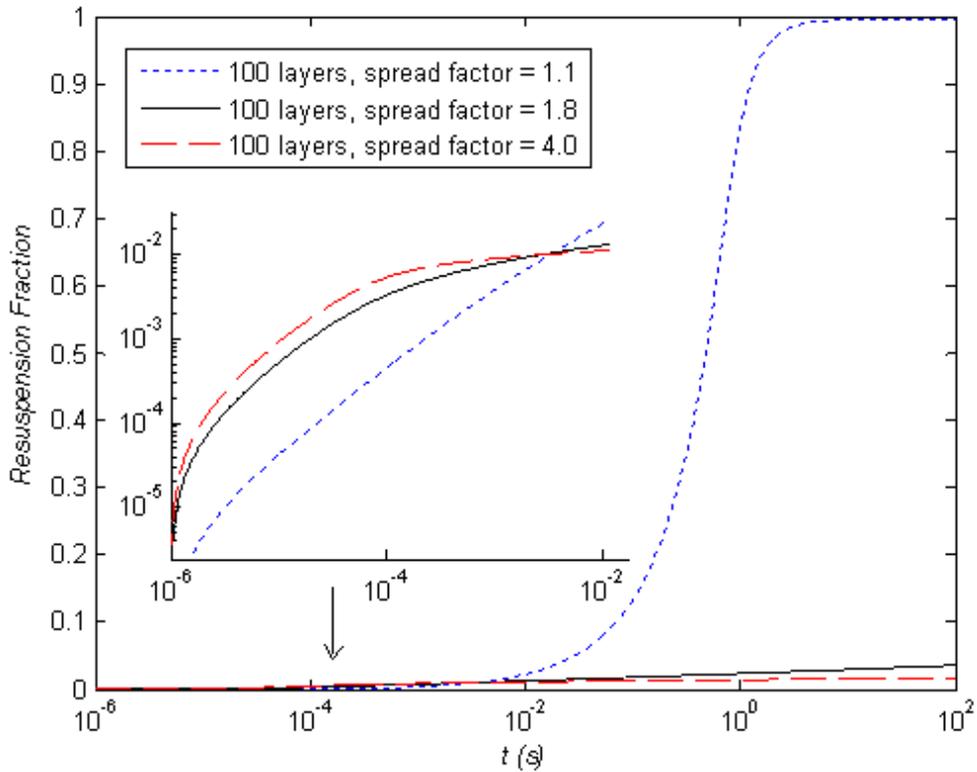

**Figure 9 -** Effect of adhesion spread factor on 100 layer deposit based on STORM test (SR11) Phase 6 conditions (see Table 2).

We note that a spread of 1.1 corresponds to a very narrow distribution of adhesive forces (a value of 1.0 corresponds to zero spread) so in this case there is only a very small amount of longer-term resuspension, most particles being resuspended over a period of $10^{-2}$s. As the number of layers increases from 1 to 100 layers, the resuspension is still relatively sharply defined but the onset of resuspension is delayed from $10^{-4}$s to $10^{-2}$s, whilst occurring over a period of say $10^{-1}$s. As the spread factor increases the difference between the 100-layer deposit and the monolayer is much more marked: for a spread of 4 after 1s, only 2% of the 100-layer deposit has resuspended compared with almost 100% for the monolayer deposit. There is in fact no sharp distinction between short- and long-term resuspension; in fact it is really all long-term resuspension in the multilayer case.

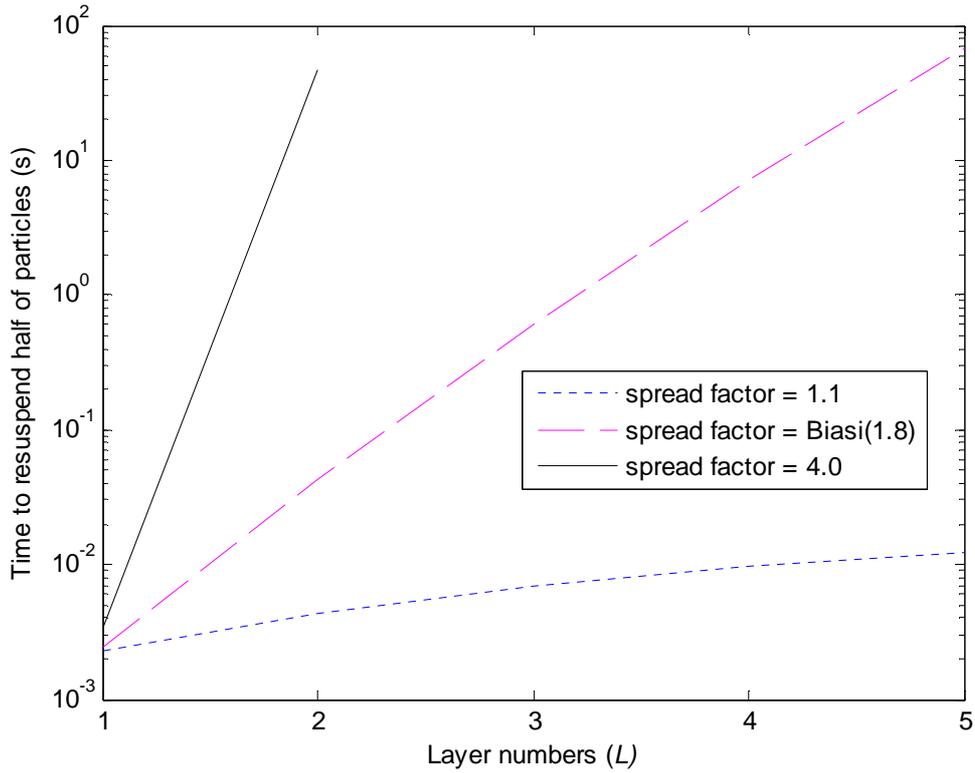

**Figure 10 -** Resuspension half-life vs. layer thickness based on STORM test (SR11) Phase 6 conditions (see Table 2).

Figure 10 shows the resuspension half-life for different spread factors. With increasing adhesive spread factor, the half resuspension time increases dramatically as the deposit becomes increasingly multilayer.

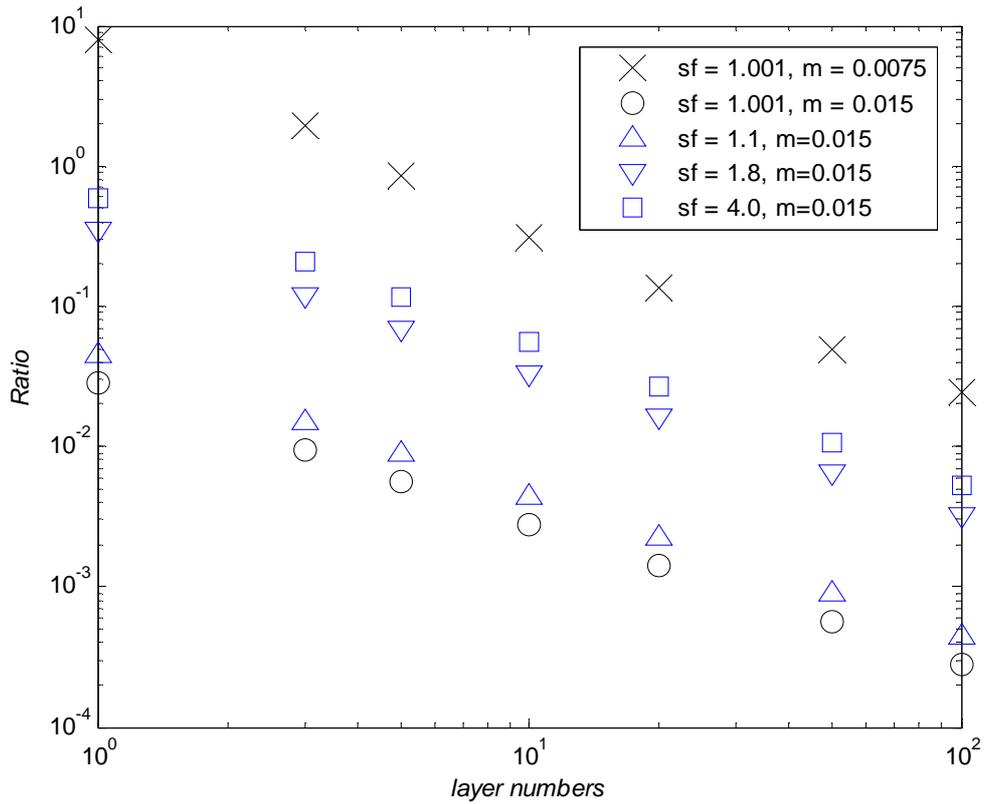

**Figure 11 -** Ratio of short-term ($<10^{-4}$ s) to long-term resuspension fraction where the long-term resuspension fraction is 1 – the short-term resuspension fraction (based on STORM test SR11 Phase 6 conditions - see Table 2) (m: geometric mean of normalized asperity radius, sf: spread factor)

Figure 11 shows the resuspension fraction short-term to long-term ratio for the flow and adhesion parameters for the STORM SR11 test shown in Table 2. The short term is considered as the period before the resuspension rate becomes approximately proportional to $1/t$. It can be observed in Figure 4 that after around $10^{-4}$ s the curves become straight lines. Therefore, short term finishes $\sim 10^{-4}$ s ($\sim 12$ in wall units). For the geometric mean in adhesion of 0.015, it is clear that as the number of layers increases, i.e., as the deposit thickness increases, the short term resuspension contributes less and less to the total resuspension since most of the particles covered are not so easily removed. For the same number of layers $L$, whilst the ratio of short-term to long-term resuspension fraction increases with increasing adhesive spread factor $\sigma'_a$, the changes in the ratio diminish as $\sigma'_a$ increases. In contrast the ratio decreases with an inverse power law decay with increasing layer thickness that is independent of the spread factor $\sigma'_a$. In the case defined by the parameters in Table 2, the short term resuspension is very small since it is a case where the mean aerodynamic force < the geometric adhesive force. By contrast Figure 11 also shows the opposite case for a geometric mean adhesion of 0.0075 when the geometric mean adhesive force < the mean aerodynamic force. In this case, the short/long term ratio increases significantly to $\sim 1$. Significantly the power law decay of the ratio with increasing layer thickness remains the same independent of adhesive spread. More precisely these observations imply that

$$\log(Ratio) = \log[f(\sigma'_a)] + slope \cdot \log(L)$$
$$Ratio = f(\sigma'_a) \cdot L^{slope}$$

[30]

where the slope is very close to -1. $f(\sigma_a')$ is a function of $\sigma_a'$ and is plotted in Figure 12 for values of $\sigma_a' > 1.5$, the value for a macroscopically smooth polished surface (Reeks and Hall, 2001). Below this value there is no transition from short to long term resuspension

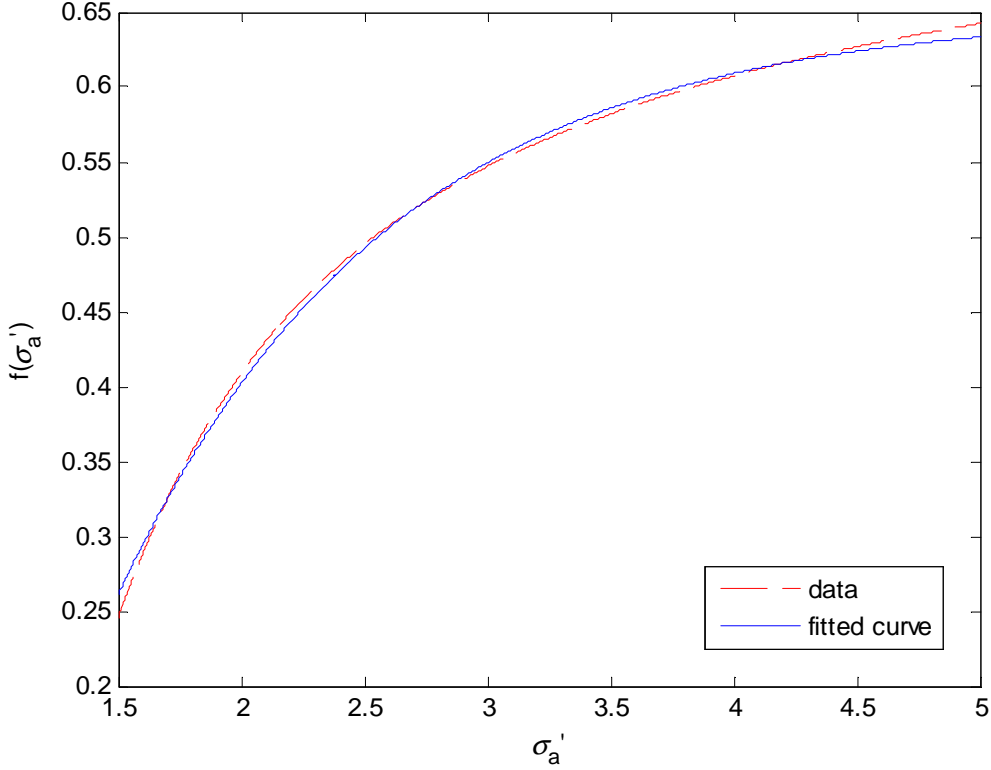

**Figure 12 -** $f(\sigma_a')$ vs. $\sigma_a'$ see Eq.(30) for the parameters given in Table 2

For The function $f$ according to Figure 12 is fitted to
$$f(\sigma_a') = -1.5\exp(-0.9\sigma_a') + 0.65 \qquad [31]$$
Note the asympotic value of 0.65 for $\sigma_a' \to \infty$.

## 4.3 Comparison of Gaussian and non-Gaussian R'n'R Models in Multilayer Resuspension

The major difference between Gaussian and non-Gaussian R'n'R models is reflected in the formula for the resuspension rate constant; we recall Reeks and Hall (2001) for the Gaussian case,

$$p = \frac{1}{2\pi}\sqrt{\frac{\langle \dot{f}^2 \rangle}{\langle f^2 \rangle}} \exp\left(-\frac{f_{dh}^2}{2\langle f^2 \rangle}\right) \bigg/ \frac{1}{2}\left[1 + erf\left(\frac{f_{dh}}{\sqrt{2\langle f^2 \rangle}}\right)\right]$$

and the non-Gaussian case Eq.[13]

$$p = B_{\dot{f}}\sqrt{\frac{\langle \dot{f}^2 \rangle}{\langle f^2 \rangle}} \frac{z_d + A_1}{A_2^2} \exp\left(-\frac{1}{2}\left(\frac{z_d + A_1}{A_2}\right)^2\right) \bigg/ \left[1 - \exp\left(-\frac{1}{2}\left(\frac{z_d + A_1}{A_2}\right)^2\right)\right]$$

with the same values of the rms coefficient $f_{rms}$ and the typical forcing frequency $\omega^+$,

$$\sqrt{\langle f^2 \rangle} = f_{rms}\langle F \rangle, \quad \sqrt{\frac{\langle \dot{f}^2 \rangle}{\langle f^2 \rangle}} = \omega^+\left(\frac{u_\tau^2}{\nu_f}\right).$$

The fractional resuspension predicted by of the multilayer model using these two R'n'R formula for the resuspension rate constant was calculated. In the comparison, the typical forcing frequency $\omega^+$ and the rms coefficient $f_{rms}$ are chosen to be the same for both Gaussian and non-Gaussian cases ($\omega^+ = 0.0413$ and $f_{rms} = 0.2$). Therefore, the difference between the curves in Figure 13 is due only to the Gaussian and non-Gaussian distributions (the distributions of aerodynamic resultant force and its derivative) used to derive the resuspension rate constant.

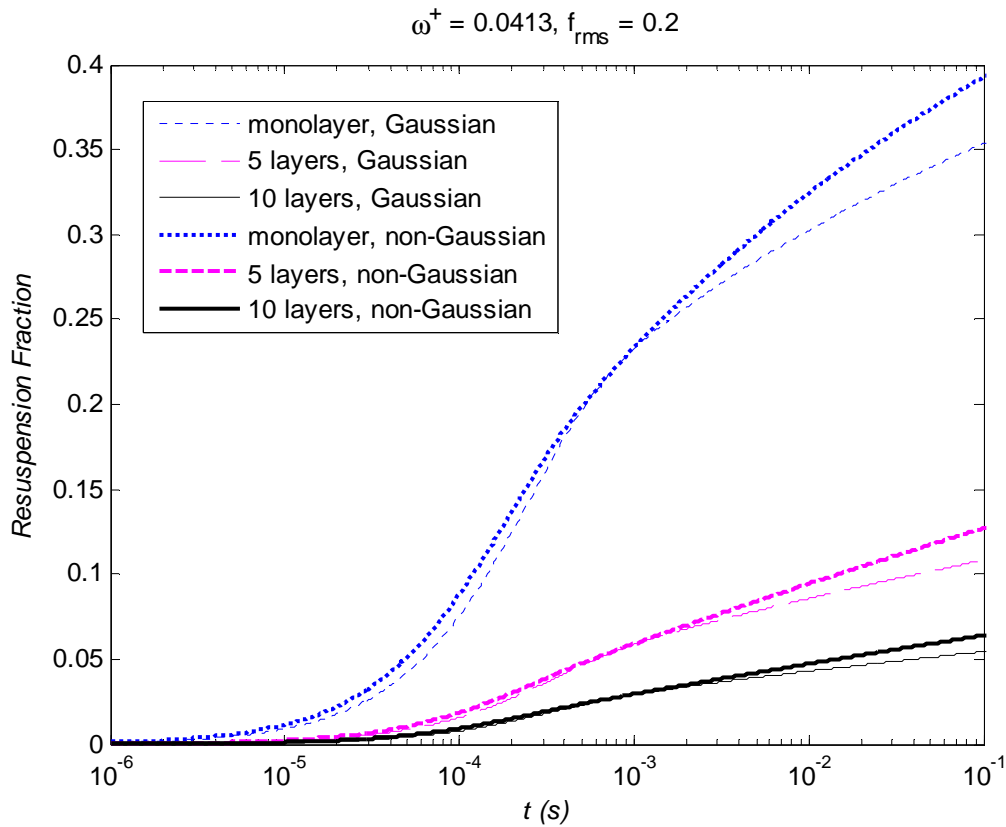

**Figure 13 -** Comparison of multilayer model with the rate constant based on Gaussian and non-Gaussian R'n'R model, STORM test SR11 Phase 6 conditions (see Table 2)

Figure 13 shows that the non-Gaussian model always results in more resuspension than the Gaussian case at all times except one point. As time increases, the difference becomes greater and greater. It is noted that the time of the contact point (around $10^{-3}$s in the plot) depends on the wall friction velocity, typical forcing frequency and the rms coefficient; it is independent of the number of layers.

## 5. Influence of Layer Coverage on Multilayer Resuspension

We now consider changes to the FY generic model effect due to coverage. We assume now that particles are not regularly spaced as in the lattice of the FY generic model but randomly spaced with the same distribution for all layers. Thus in any given layer apart from the top layer when no particles are removed from the layer above, an area corresponding to the particles projected cross section area is revealed within which particles are exposed to the flow. Here a new parameter $\alpha_i$ (the coverage coefficient) is introduced to describe the number ratio of particles exposed to the flow in a given layer ($i + 1$) due to removal of particles from the layer above ($i$), namely

$$a_i = f_{Ai} \frac{\bar{n}_{i+1}}{\bar{n}_i} \quad [32]$$

where $f_{Ai}$ is the particle occupied-area fraction (the fraction of particles projected area in horizontal plane parallel to the wall) in layer $i$ and $\bar{n}_i$ is the initial average number of particles in the $i$th layer. In other words, all the particles in layer $i$ would therefore cover $f_{Ai}\bar{n}_{i+1}$ particles in layer $i + 1$. It is noted that the average number of particles in each layer is the same, therefore $\bar{n}_{i+1}/\bar{n}_i = 1$. Also it is assumed that the coverage coefficient is the same for all layers (use $\alpha$ instead of $\alpha_i$). The particle occupied-area fraction is defined as the ratio of the total (remaining) particle cross-section area to the whole area of the layer.

$$f_{Ai} = \frac{\bar{A}_{pi}\bar{n}_i}{A_i} \quad [33]$$

where $\bar{A}_{pi}$ is the average particle projected area (for a spherical particle, this is $\pi \bar{r}_i^2$ and $\bar{r}_i$ is the average particle radius in layer $i$) and $A_i$ is the total surface area of layer $i$.

The average number of particles in the $i$th layer can be determined as

$$\bar{n}_i = \frac{V_i f_{Vi}}{\bar{V}_{pi}} \quad [34]$$

where $V_i$ is the total volume of layer $i$. $f_{Vi}$ is the particle volume fraction in the $i$th layer and $\bar{V}_{pi}$ is the average single particle volume in the $i$th layer. The volume fraction is related by definition to the porosity, namely

$$f_{Vi} = 1 - \varsigma_i \quad [35]$$

It is noted that, strictly, this porosity ought to be different for different layers. Here, it is assumed that the porosities for all the layers are the same (henceforth we use $\varsigma$).

The average thickness of the $i$th layer is assumed to be the average diameter of particles in that layer. Therefore, the particle occupied area fraction is given by

$$f_{Ai} = \frac{\bar{A}_{pi}\bar{n}_i}{A_i} = \frac{\bar{A}_{pi}V_i f_{Vi}}{A_i \bar{V}_{pi}} = \frac{\bar{A}_{pi} 2\bar{r}_i}{\bar{V}_{pi}}(1-\varsigma) \quad [36]$$

As stated above, the average number of particles in each layer is the same so the coverage coefficient can therefore be derived as

$$a = f_{Ai}\frac{\bar{n}_{i+1}}{\bar{n}_i} = \frac{\bar{A}_{pi} 2\bar{r}_i}{\bar{V}_{pi}}(1-\varsigma) \quad [37]$$

For a deposit composed of spherical particles,

$$f_{Ai} = \frac{\pi \bar{r}_i^2 \bar{n}_i}{A_i} = \frac{\pi \bar{r}_i^2 2A_i \bar{r}_i f_{Vi}}{A_i \,{}^4\!/_3 \pi \bar{r}_i^3} = \frac{3}{2}f_{Vi} = \frac{3}{2}(1-\varsigma) \quad [38]$$

Therefore, the coverage coefficient for the deposit of spherical particles according to Eq.[32] and Eq.[38] is given by

$$\alpha = \frac{3}{2}(1-\varsigma) \qquad [39]$$

Note: $\alpha$ varies from 0 to 1.5. $\alpha = 1$ means that if one particle is removed from the layer above only one particle in the layer below will be exposed to the flow. Friess and Yadigaroglu (2002) demonstrated that in STORM SR11 experiment conditions, their deposition and resuspension model with the porosity between 0.62 and 0.71 gave the closest results to the experimental data. Therefore, putting these values into Eq.[39], the resulting value of $\alpha$ in STORM SR11 experiment conditions is calculated to be from 0.44 to 0.57.

$\alpha$ is the number ratio of particles exposed to the flow of the layer below to the current layer and than it can be derived as the rate ratio since the exposure rate is approximately the number of particles multiplied by the rate constant. Thus,

$$a = f_{Ai} \frac{\overline{n}_{i+1}}{\overline{n}_i} \qquad f_{Ai} \frac{\dot{\overline{n}}_{i+1}}{\dot{\overline{n}}_i} \qquad [40]$$

Therefore, Eq.[23] is redefined by including the exposure-rate ratio in the source term.

$$\frac{\partial n_1(r'_a,t)}{\partial t} = -p(r'_a)n_1(r'_a,t)$$

$$\frac{\partial n_i(r'_a,t)}{\partial t} = -p(r'_a)n_i(r'_a,t) + \alpha \cdot \varphi(r'_a)\int_0^\infty p(\tilde{r}'_a)n_{i-1}(\tilde{r}'_a,t)d\tilde{r}'_a \quad (i \geq 2) \qquad [41]$$

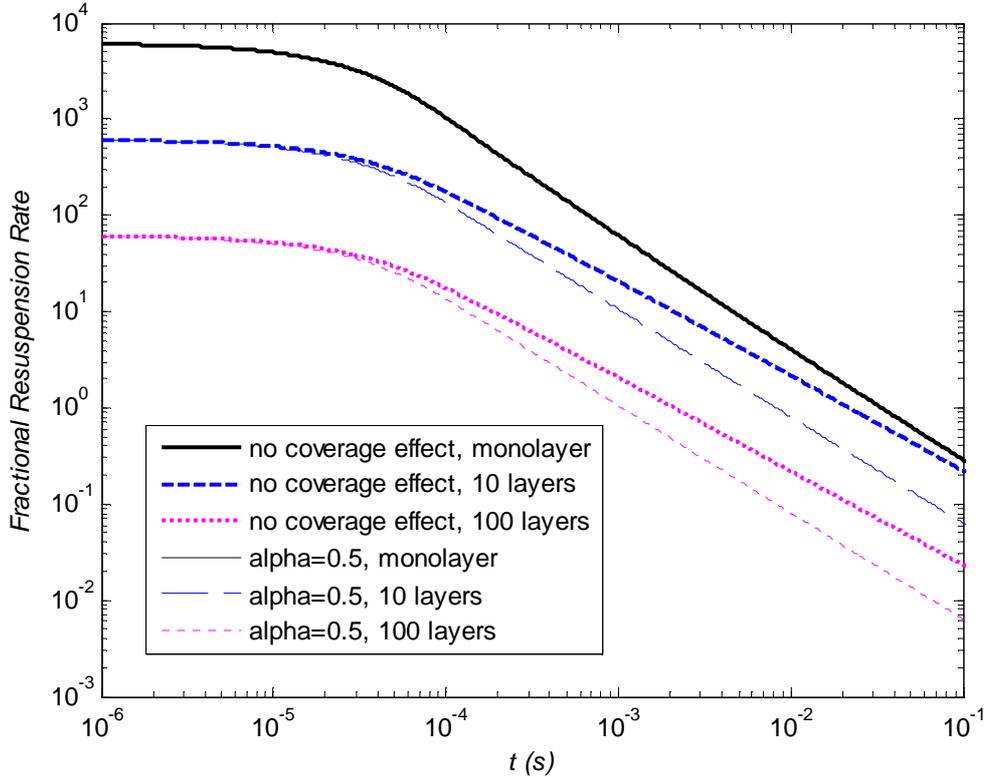

**Figure 14** - Effect of exposure-rate ratio on fractional resuspension rate, based on STORM test SR11 Phase 6 conditions (see Table 2)

Figure 14 and Figure 15 show the resuspension-rate and fraction comparison between the initial multilayer model and the model including the coverage coefficient. As one can observe from the resuspension-rate comparison, the monolayer case is unaffected by the coverage coefficient as expected while for the multilayer case (10 and 100 layers) the resuspension rate with no coverage effect (with $α = 1$) is very close to the resuspension with $α = 0.5$ in the short term (i.e., $<10^{-4}$s) exposure time (since this is mainly due to resuspension of the top layers where the reduced coverage effect has hardly any impact: only in the long-term (i.e., $>10^{-4}$s) does the coverage effect start to dominate by dramatically reducing the resuspension rate. This can also be observed in Figure 15 which is the comparison of resuspension fraction as a function of time. A coverage coefficient $α = 0.5$ reduces the fractional amount by around half after 100s. However, this coverage coefficient is determined by the porosity of each layer and in reality the porosity for each layer is different and this porosity varies as time goes on. In the next section, this ratio constant will be replaced with a particle size distribution which will be more physically correct.

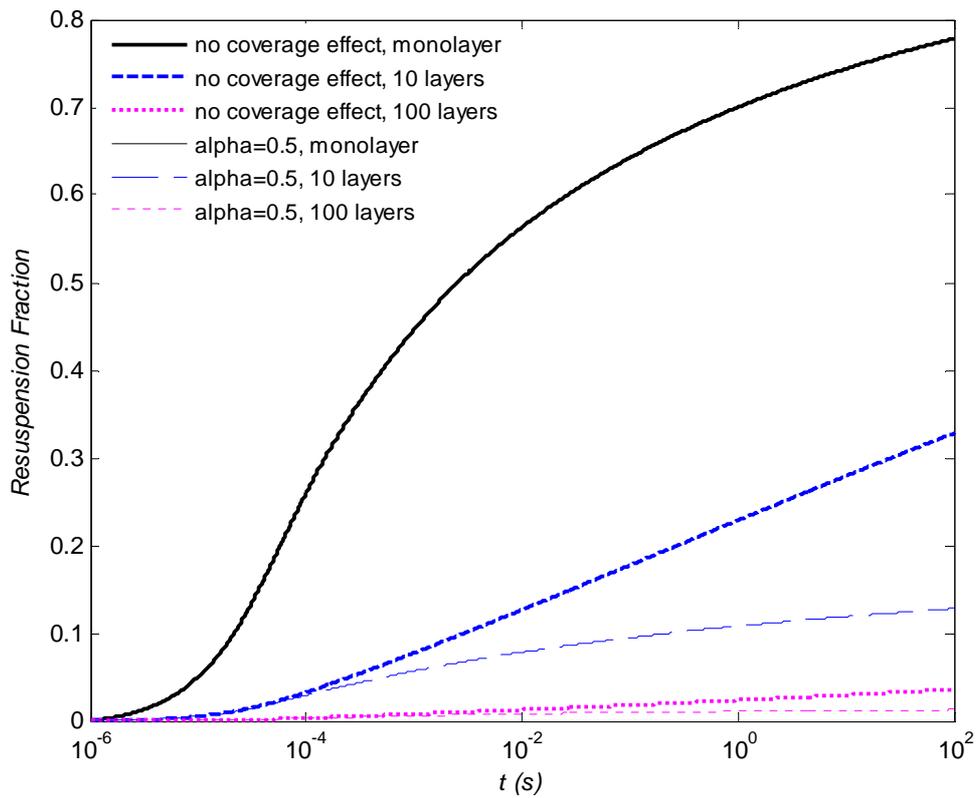

**Figure 15 -** Effect of exposure-rate ratio on resuspension fraction, based on STORM test SR11 Phase 6 conditions (see Table 2)

## 6. Multilayer Resuspension with Particle Size Distribution in Each Layer

In order to study the behaviour of deposit structure in particle resuspension, the clustering of particles within a given layer and between layers (the clustering effect) is an important consideration. However, little information is to be found on the cluster and deposit structure in a bed of particles. In this section, a simple addition is introduced to the set of multilayer equations to include the influence of particle size distribution in the initial multilayer model where the clusters are considered simply as different size particles. The interaction between the particles in the same layer is not explicitly considered, but the coverage effect from the layer above to the layer below will be included.

The distribution $\psi(r)$ of the particle radius $r$ is considered as log-normal with geometric mean $\bar{r}$ and geometric standard deviation $\sigma_r$, i.e.

$$\psi(r) = \frac{1}{\sqrt{2\pi}} \frac{1}{r \ln \sigma_r} \exp\left(-\frac{[\ln(r/\bar{r})]^2}{2(\ln \sigma_r)^2}\right) \qquad [42]$$

Since the resuspension rate constant $p$ and exposure pdf $n$ is a function of particle size, then the formula for the pdf of exposed particles (Eq.[23]) becomes

$$\frac{\partial n_1(r, r_a', t)}{\partial t} = -p(r, r_a') n_1(r, r_a', t)$$

$$\frac{\partial n_i(r, r_a', t)}{\partial t} = -p(r, r_a') n_i(r, r_a', t) + \psi(r)\varphi(r_a') \int_r^\infty \int_0^\infty p(\tilde{r}, \tilde{r}_a') n_{i-1}(\tilde{r}, \tilde{r}_a', t) d\tilde{r}_a' d\tilde{r} \quad (i \geq 2) \qquad [43]$$

Note: the integration in the source term of the second equation for particle size distribution is taken from the current particle radius $r$ to infinity. In other words, in order to expose the current particle of radius $r$ to the flow, the radius of the particle from the previous layer must be larger than the current particle radius.

Each deposit layer is assumed as the formation of a distribution of spherical particles as shown,

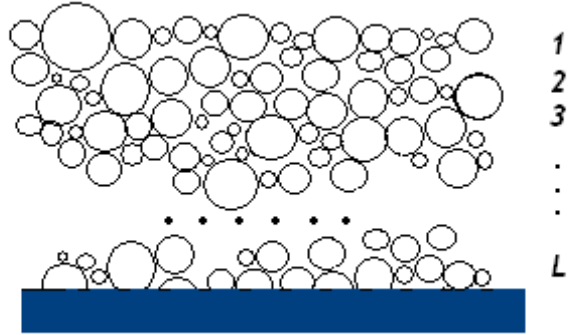

**Figure 16** - Geometry of the multilayer system with polydisperse particles

The resuspension rate in the $i$th layer is thus defined as

$$\Lambda_i(t) = \int_0^\infty \int_0^\infty p(r, r_a') n_i(r, r_a', t) dr_a' dr \qquad [44]$$

The roughness condition for all the particles is assumed to be the same. The same distribution of normalized asperity radius ($\varphi(r_a')$) is used for all the particles in the domain. Layer thickness is determined by the mean particle diameter and they are assumed to be the same for every layer.

In the calculation below, the parameters from the STORM SR11 experiment Phase 6 conditions will be applied together with Biasi's correlation for the parameters of the adhesion distribution.

| Average radius ($\mu m$) | Geometric standard deviation of radius | Fluid density ($kg.m^{-3}$) | Fluid kinematic viscosity ($m^2.s^{-1}$) | Wall friction velocity ($m.s^{-1}$) | Surface energy ($J.m^{-2}$) |
|---|---|---|---|---|---|
| 0.227 | 1.7 | 0.5730 | $5.2653 \times 10^{-5}$ | 6.249 | 0.5 |

**Table 3 -** Parameters in STORM SR11 Phase 6 for polydisperse model

The average particle size and geometric standard deviation are taken from those used in an analysis with the SOPHAEROS code (Cousin *et al.*, 2008).

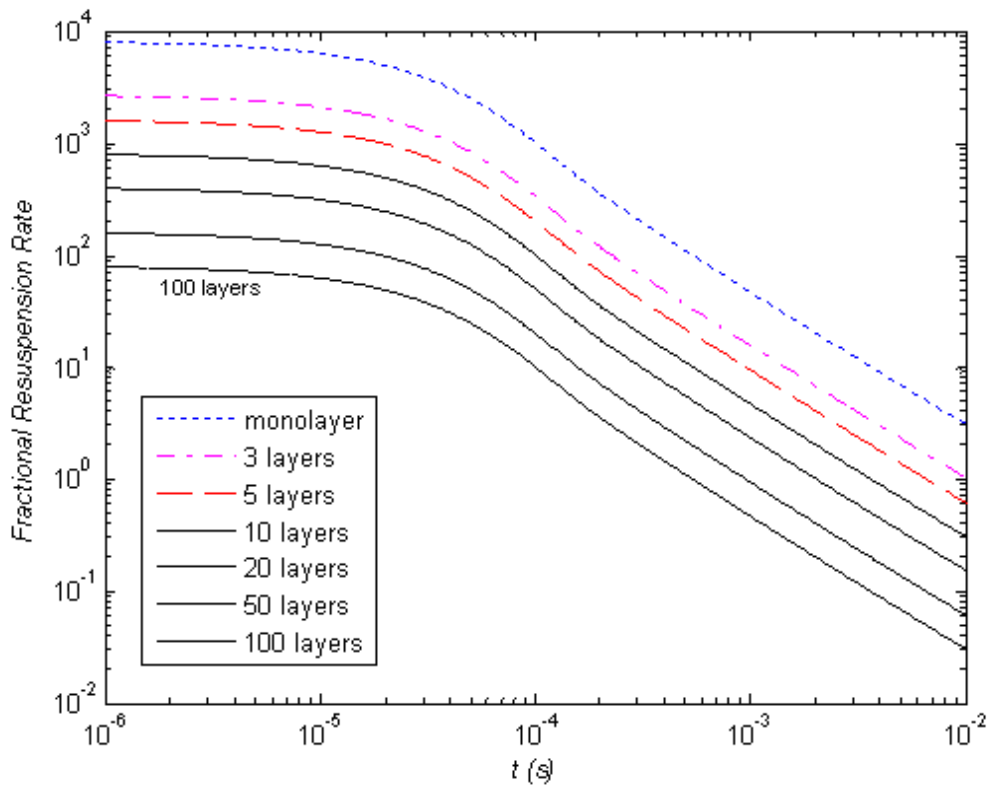

**Figure 17 -** Fractional resuspension rate vs. time for model with polydisperse particles, STORM test SR11 Phase 6 conditions (see Table 3)

The fractional resuspension rate (calculated by Eq.[28]) for the polydisperse particles is shown in Figure 17. Compared to the results for monodisperse particles (Figure 14), the long-term resuspension rate in this case substantially decreases when the number of layers increases. Below, Figure 18 compares the monodisperse and polydisperse resuspension results for aerosols having the same geometric mean size (Eq.[42]). Interpreting Figure 18, the comparison of the fractional resuspension rates shows that the fractional resuspension of the polydisperse particles in the short term is greater than that for the monodisperse particles due to the fact the resuspension rate is biased towards larger-size particles. After the initial resuspension and the larger particles (> the geometric mean) have been removed, the remaining polydispersé particles are biased towards particle sizes < the geometric mean which are harder to remove than those with the geometric mean size. Note that the long-term resuspension rate of the polydisperse case is several orders of magnitude less than that of the monodisperse case.

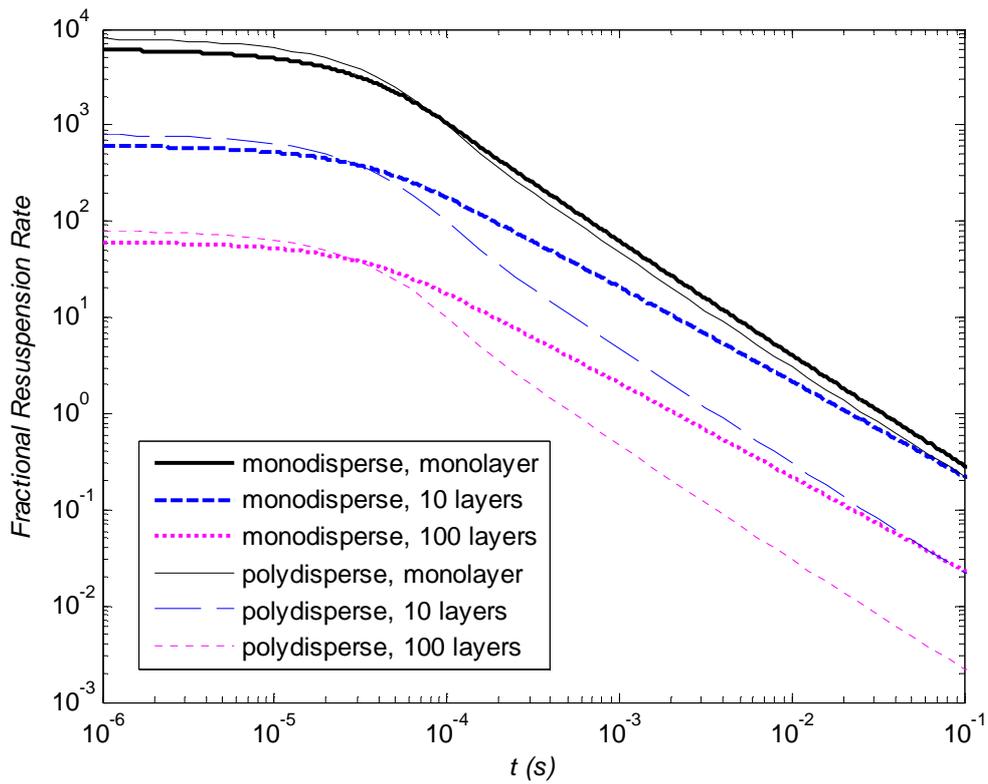

**Figure 18 -** Comparison of fractional resuspension rate vs. time for two multilayer models, STORM test SR11 Phase 6 conditions (see Table 3)

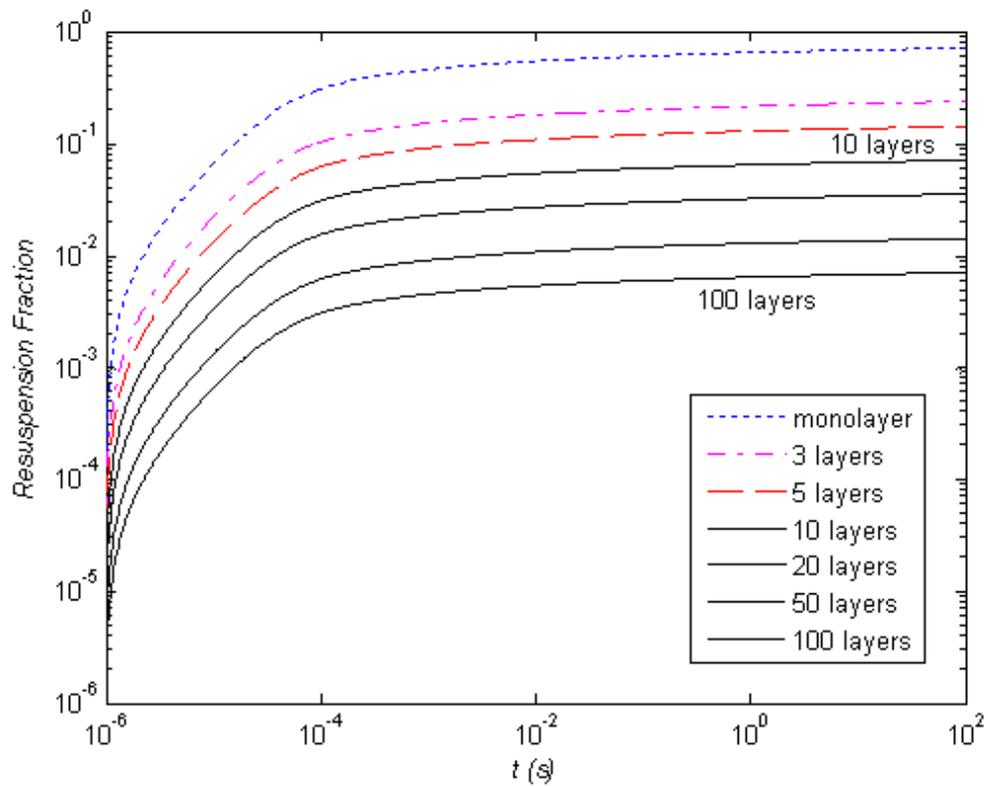

**Figure 19 -** Resuspension fraction vs. time for polydisperse particles, STORM test SR11 Phase 6 conditions (see Table 3)

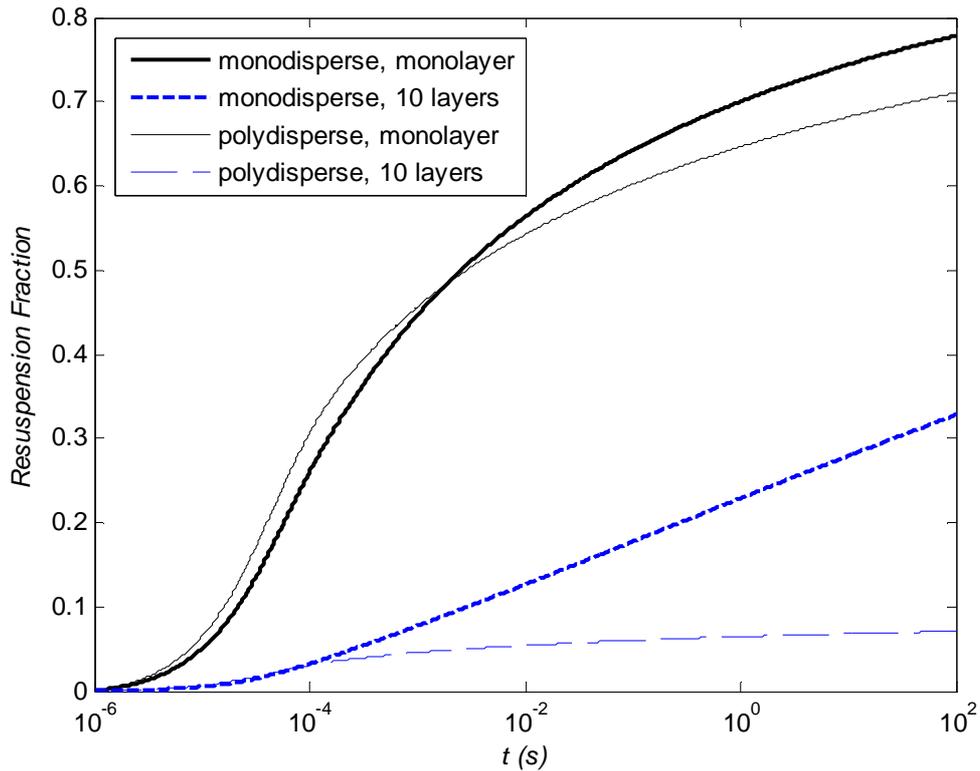

**Figure 20 -** Resuspension fraction vs. time for polydisperse and monodisperse particles with the same geometric mean size, STORM test SR11 Phase 6 conditions (see Table 3)

Figure 19 shows the resuspension fraction for different numbers of layers for the polydisperse case. It is noted that as the number of layers increases to more than 10, less than 10% of the deposit is removed after 100s. It also shows that the resuspension fraction for each case is exponentially proportional to the time in the long-term after $10^{-4}$s. The comparison of the resuspension fraction for the polydisperse particles and the monodisperse particles with the same geometric mean size and is shown in Figure 20. This confirms what was said above that, compared to the monodisperse resuspended fraction, the polydisperse case gives slightly more resuspension than the monodisperse case in the short initial period since a population of larger-than-average-size particles is considered in the polydisperse model and these are easier to remove than smaller ones. After that initial period, the resuspension fraction of polydisperse particles becomes less and less compared to that of the monodisperse particles since the remaining polydisperse particles are smaller and increasingly harder to remove.

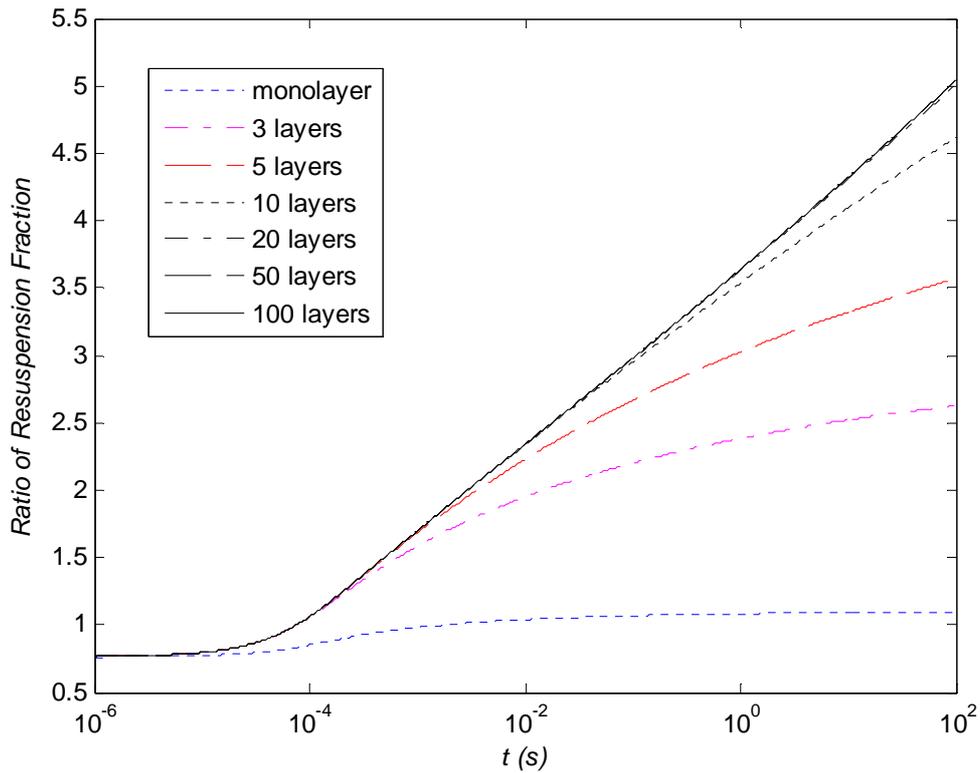

**Figure 21 -** Ratio of resuspension fraction between initial monodisperse model and model with polydisperse particles, STORM test SR11 Phase 6 conditions (see Table 3)

Figure 21 shows the ratio of the resuspension fraction of the monodisperse particles with that of the polydisperse particles. All the curves start with similar values due to the resuspension of the top layers; also in the short term ($<10^{-4}$s) the resuspension of the polydisperse particles is greater than that of the monodisperse particles since the 50% of the polydisperse particles with sizes > that of the monodisperse particles are easier to resuspend. As the number of layers increases the difference between the monodisperse and polydisperse particles becomes significant. For 10 layers, the resuspension fraction of the monodisperse particles after 100s is ~4.5 times that of polydisperse particles. And for the 100 layers case, the ratio increases to around 5 after 100s. This is all consistent with the Figures provided for the resuspension rate comparison.

It is noted here that in the calculations of the multilayer polydisperse particles, Biasi's correlation of the parameters for the distribution of adhesive forces was used for each particle in the size distribution; so the geometric mean of normalized asperity radius and adhesive spread factor is different for every particle size.

To examine the effect of spread factor, the value of the geometric mean normalized adhesive radius is now fixed (a value of 0.015 is used) and we consider the influence of spread factors of 1.1 (a very narrow spread), 1.817 (Biasi's original value) and 4.0 with the same size distribution as before in Figure 19.

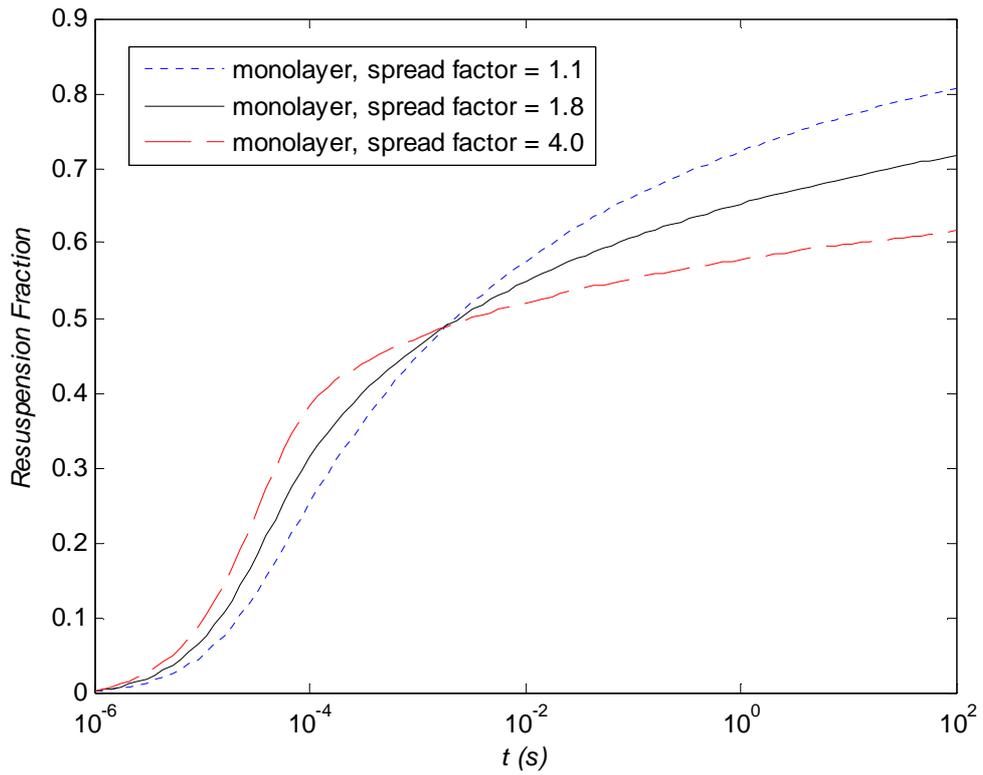

**Figure 22 -** Effect of adhesive spread factor on monolayer resuspension for polydisperse particles, STORM test SR11 Phase 6 conditions (see Table 3)

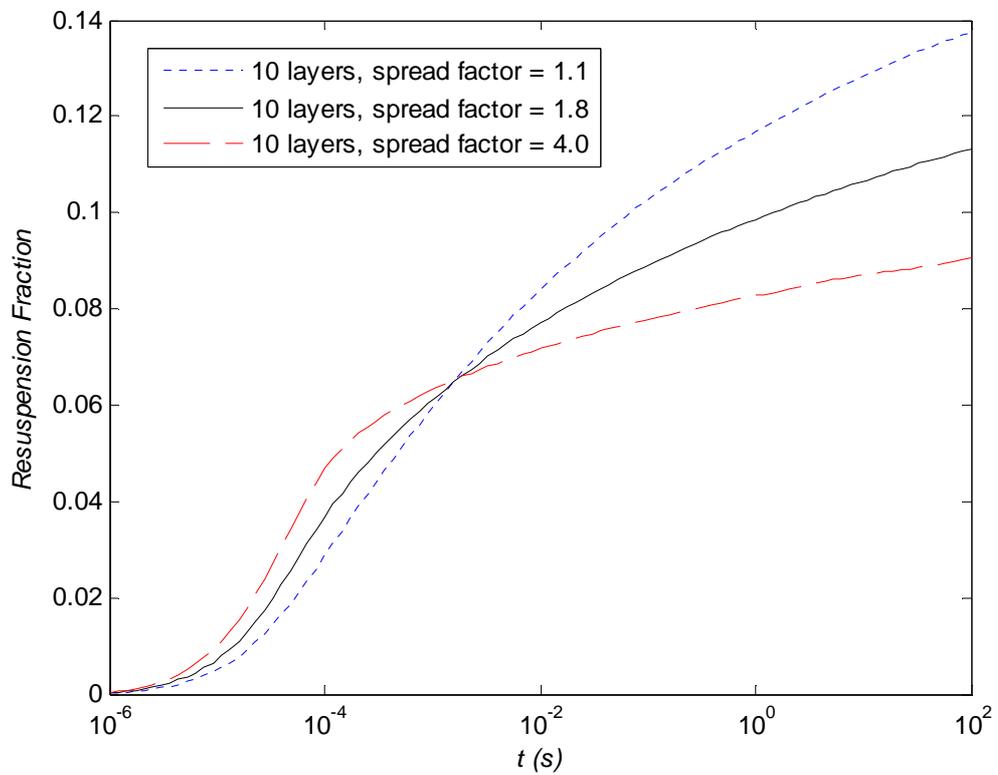

**Figure 23 -** Effect of adhesive spread factor on resuspension of polydisperse particles in a 10-10-layer multilayer deposit, STORM test SR11 Phase 6 conditions (see Table 3)

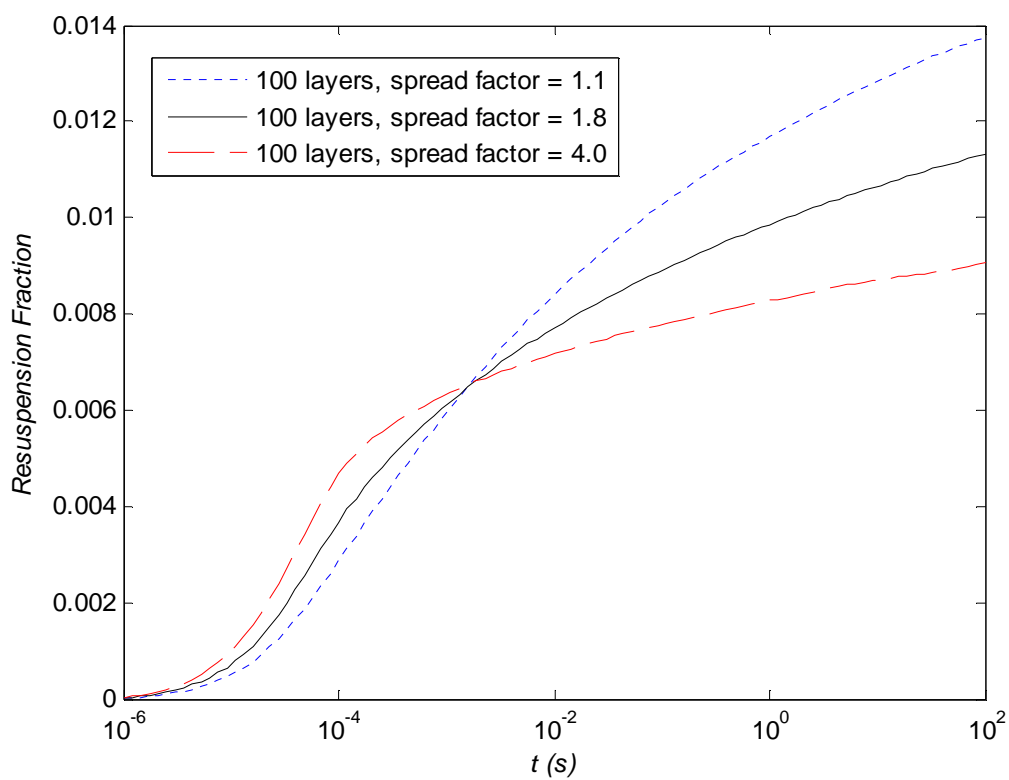

**Figure 24 -** Effect of adhesive spread factor on resuspension of polydisperse particles in a 100-layer multilayer deposit, STORM test SR11 Phase 6 conditions (see Table 3)

Figures 22, 23 and 24 show the effect on the fraction resuspended of spread factors for a range of layer thicknesses in the deposit. It is observed in Figure 22 that the monolayer model with larger spread initially gives more resuspension due to the fact that there is more large size particles involved in the distribution having lower adherence and these are easier to remove. After a certain period ($10^{-3}$s), the model with the larger adhesive spread predicts less and less resuspension because the particles left are much harder to remove. However, as the number of layers increases the difference between the resuspension for different spread factors becomes less and less because the coverage effect is increasingly influential in making particles are harder to remove in the thicker deposit.

It is noted that in Figures 7, 8 and 9, the curves for the small adhesive spread factor (1.1) are quite different to the others with larger spread factors even for large layer numbers (the curve for small spread factor being closer to a step function). However, this phenomenon is not reproduced here. The reason for this is because when a particle size distribution is considered, there exists a large probability of small particles (radius smaller than the average radius). Therefore, no matter how small is the range of adhesive force is, there is still a large fraction of particles needing to be removed over a much longer period.

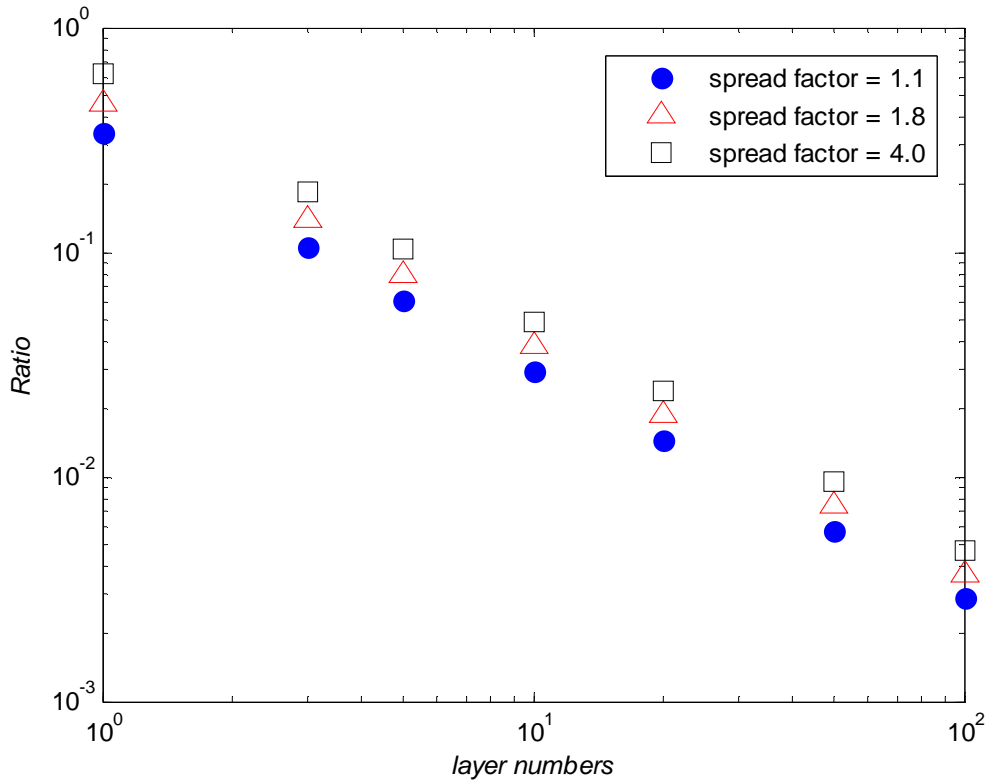

**Figure 25 -** Ratio of resuspension fraction between short term (<$10^{-4}$ s) and long-term (note that the long-term resuspension is 1 – the short-term resuspension) based on STORM test SR11 Phase 6 conditions (see Table 3)

Figure 25 shows the ratio of short-term to long-term resuspension fraction. The short term is considered to be the period before the resuspension rate becomes approximately proportional to 1/t. It is observed in Figure 17 that after around $10^{-4}$s the curves become straight lines. Therefore, short term finishes around $10^{-4}$s. As the number of layers increases, in other words the deposit becomes thicker and thicker, the short term resuspension contributes less and less to the total resuspension because most of the particles were covered and not easy removed. Also, for the large spread factor (i.e., 4) the ratio of short-term to long-term resuspension fraction is always higher than the model with small spread factor though the difference is less significant when the number of layers is very large. The reason for this is the wide range of adhesive force where the particles with small adhesive force being more easily removed in the short term and this affects the layers on the top more than the layers below. Also, Figure 25 indicates that the slopes are the same for the range of spread factors. It is noted that the slopes are close to -1.1.

## 7. Validation of Multilayer Models

In this final section, model predictions are compared with the results of two experiments: the STORM SR11 test (Castelo *et al.*, 1999) and the BISE experiment (Alloul-Marmor, 2002). The former is probably the best experiment for multilayer deposit resuspension in the nuclear safety area and the latter is the latest experiment for multilayer resuspension for which the experimental data have been compared with R'n'R model predictions.

### 7.1 STORM Test SR11

The STORM (Simplified Test of Resuspension Mechanism) experiements were carried out by the Joint Research Centre of the European Commission (EC/JRC) at Ispra, Italy. The aerosol used throughout the tests was composed of tin oxide ($SnO_2$) particles produced by a plasma torch. In all, 13 tests were undertaken, the first 8 tests only for aerosol deposition and the rest of the tests included a supplementary resuspension phase. The 5 resuspension tests were identified as SR09, SR10, SR11, SR12 and SR13 (Bujan *et al.*, 2008). SR11 is also the basis of an International Standard Problem (ISP-40, in fact; Castelo *et al.*, 1999). It took place in April 1997 and included two distinct phases, the first concentrating on aerosol deposition mostly by thermophoresis and eddy impaction (turbulence) and the second on aerosol resuspension under a stepwise increasing gas flow. We note that the dominant deposition mechanisms should in the STORM conditions not have favoured deposition of any particular size in the case of thermophoresis (i.e., a polydisperse deposit) but that turbulent deposition must have strongly favoured large-particle deposition (much reduced polydispersity).

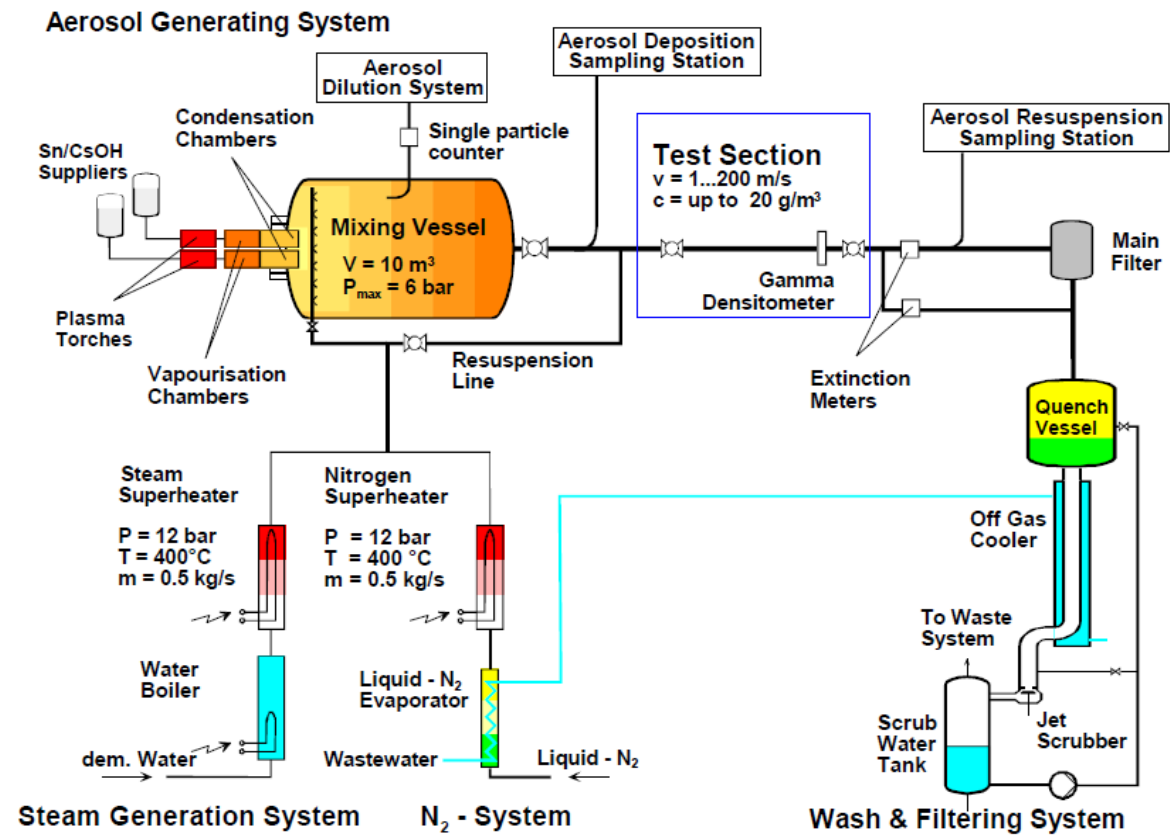

**Figure 26** - The STORM facility (Castelo *et al.*, 1999)

The STORM test facility is shown in Figure 26. The test section (in the blue box) was a 5.0055m long straight pipe with 63 mm internal diameter. In the deposition phase, the carrier gas and aerosols passed through the mixing vessel followed by a straight pipe into the test section and then finally to the wash and filtering system. In the resuspension phase, the clean gas was injected through the resuspension line directly into the test section and the resuspended $SnO_2$ aerosols were collected on the main filter before the gas goes through the wash and filtering system (Castelo *et al.*, 1999). The resuspension phase was divided into six steps of increasing gas velocity; the carrier gas was pure nitrogen ($N_2$) at 370 ºC; the flow rates, fluid mean velocity, fluid density and kinematic viscosity for each step are given in the Table 4 below.

| Step | Mass flow rate (kg/s) | Fluid mean velocity (m/s) | Fluid density (kg/m$^3$) | Fluid kinematic viscosity (m$^2$/s) | Wall friction velocity (m/s) |
|---|---|---|---|---|---|
| 1 | 0.102 | 62.01 | 0.4422 | 6.0697 x 10$^{-5}$ | 2.773 |
| 2 | 0.126 | 76.87 | 0.5425 | 5.5521 x 10$^{-5}$ | 3.438 |
| 3 | 0.152 | 93.17 | 0.5480 | 5.5000 x 10$^{-5}$ | 4.167 |
| 4 | 0.175 | 107.78 | 0.5566 | 5.4204 x 10$^{-5}$ | 4.820 |
| 5 | 0.199 | 123.28 | 0.5647 | 5.3427 x 10$^{-5}$ | 5.513 |
| 6 | 0.224 | 139.74 | 0.5730 | 5.2653 x 10$^{-5}$ | 6.249 |

**Table 4 -** Conditions of STORM test SR11

While the mass flow rate was measured in STORM test SR11, the fluid mean velocity, density and kinematic viscosity are obtained from the SOPHAEROS code (Cousin *et al.*, 2008); SOPHAEROS is the module of the European integral code ASTEC (Accident Source Term Evaluation Code) that has been used by Bujan *et al.* (2008) and others to analyse the experimental results of the STORM resuspension tests. The wall friction velocity is calculated as

$$u_\tau = \sqrt{\frac{\tau_w}{\rho_f}}$$

where $\tau_w$ is the wall shear stress, whose value can be obtained from a 1-D thermal-hydraulic system code simulation, using the formula

$$\tau_w = \frac{f_{fric}}{8}\rho_f V^2 \qquad [45]$$

where $V$ is the fluid mean velocity and $f_{fric}$ is the friction factor which in the STORM case is approximately 0.016 (Komen, 2007).

The size of particles collected after deposition in the STORM test can be reproduced by a log-normal distribution and in the SOPHAEROS code the geometric mean diameter (GMD) is given as 0.454μm and the geometric standard deviation 1.7.

The $SnO_2$ aerosol was generated during the deposition phase. The total mass of the aerosol deposited in the test pipe was estimated to be 0.162kg. The time and mass at the end of each resuspension step and the resuspension fraction are shown in the Table 5 below.

| Step | Time at the end of each step (s) | Mass at the end of each step (kg) | Fraction resuspended |
|---|---|---|---|
| 1 | 720 | 0.156 | 0.037 |
| 2 | 2280 | 0.151 | 0.068 |
| 3 | 3300 | 0.124 | 0.235 |
| 4 | 4380 | 0.096 | 0.407 |
| 5 | 5400 | 0.070 | 0.568 |
| 6 | 5820 | 0.042 | 0.741 |

**Table 5 -** Resuspension result of STORM SR11 test

The surface energy of $SnO_2$ according to Yuan *et al.* (2002) varies from 0.47 J/m$^2$ to 0.51 J/m$^2$ at 370 ºC. Here 0.5 J/m$^2$ is used for the model calculation.

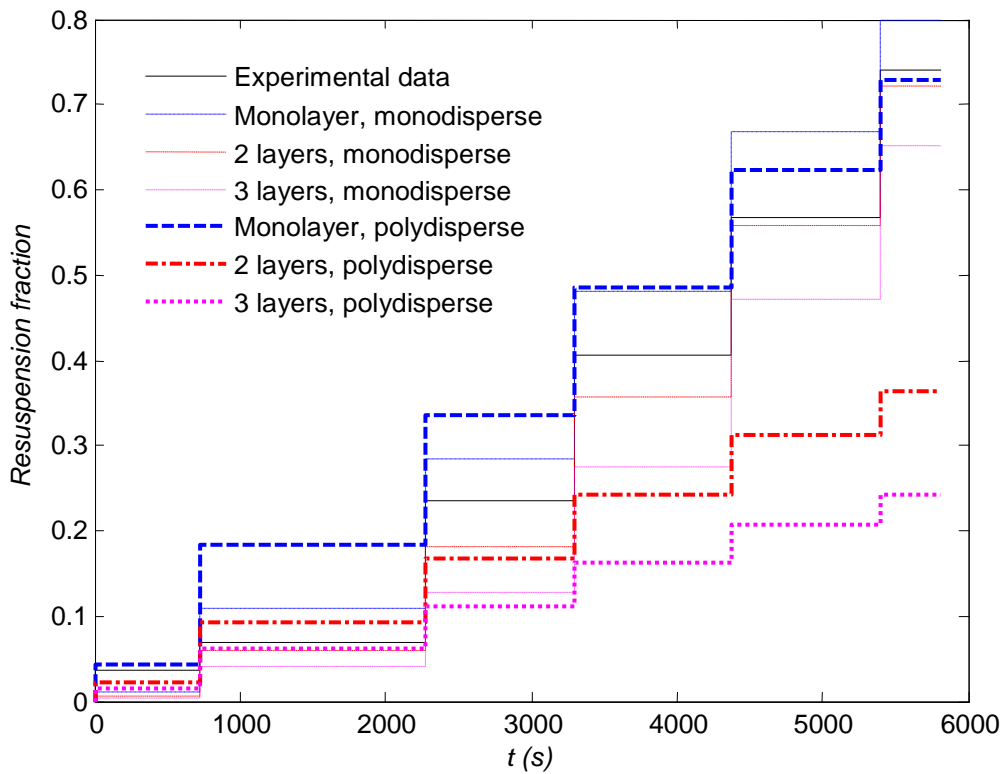

**Figure 27 -** STORM SR11 test and model results comparison (using Biasi's correlation [Eq.7] for the adhesive forces)

For the purposes of comparison, two multilayer models were used to calculate the fraction of particles resuspended at the end of each time step: the monodisperse multilayer model (referred to as the monodisperse case) and the polydisperse multilayer model (referred to as the polydisperse case). The values of the parameters used in the model predictions are those given in Tables 4 and 5. Biasi's correlation for the adhesive force distribution according to each particle size was used to obtain the results shown in Figure 27.

Figure 27 shows the model comparisons of the resuspension fraction with the STORM test SR11 results. It is observed that for the monolayer resuspension both polydisperse and monodisperse cases give more resuspension than the experimental data. However, the polydisperse case gives results very close to the experimental data in the first and final steps. Comparing all cases, the monodisperse case with 2 layers gives the best results. This would imply that the size-selective deposition of turbulent deposition significantly reduced the polydispersity of the deposit.

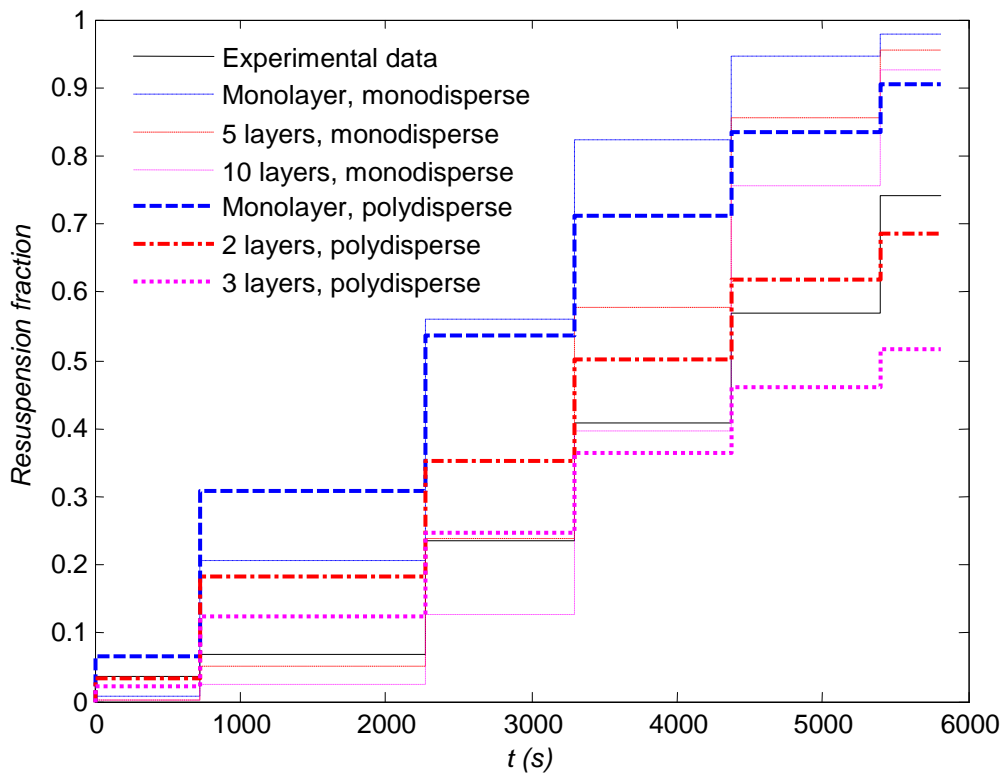

**Figure 28** - STORM SR11 test and model results comparison (geometric mean (reduction in adhesion) = 0.01, adhesive spread factor = 1.5)

However, it is recalled that Biasi's correlation for the adhesive forces is based on the original R'n'R model which is an isolated particle model for resuspension. Therefore, it is strictly not suited for the multilayer deposit case. In order to check the sensitivity of the resuspension to the values of the adhesion parameters, a geometric mean (reduction in adhesion) was chosen to be 0.01 according to Hall's experiment (Reeks & Hall, 2001) with two different spread factors (1.5 and 4.0) used for comparison. The small spread case is shown in Figure 28; one can observe that, this time, the polydisperse case with 3 layers is in much closer agreement with the experimental measurements than the monodisperse case.

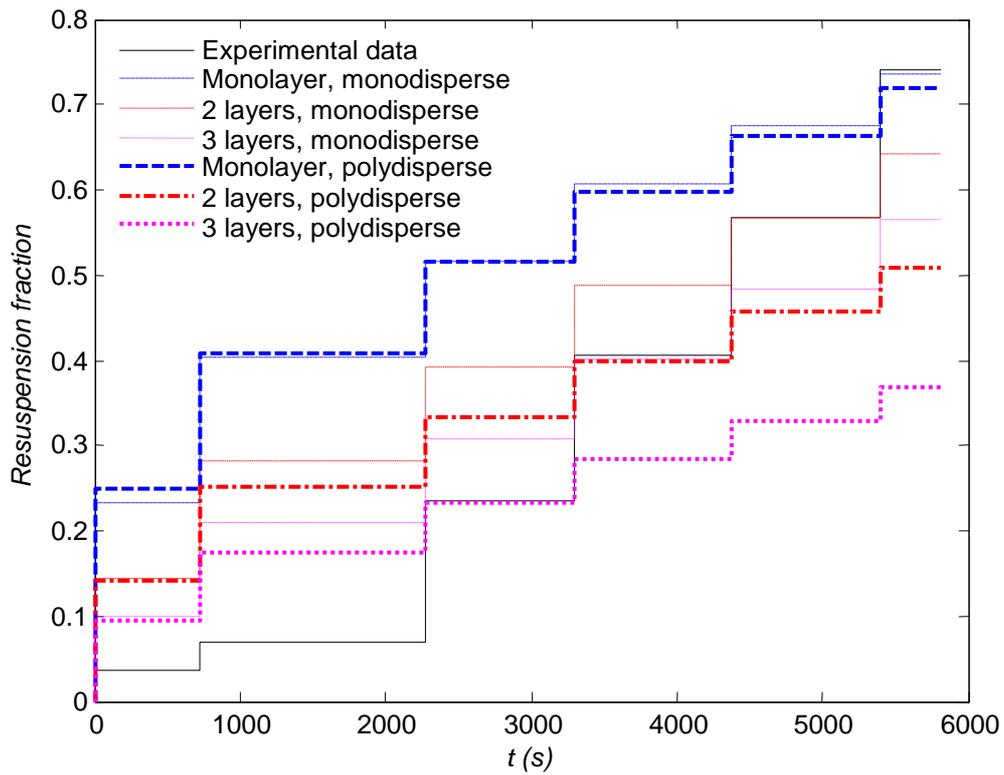

**Figure 29 -** STORM SR11 test and model results comparison (geometric mean (reduction = 0.01, adhesive spread factor = 4.0)

In the case of the resuspension for the large spread factor (4.0), Figure 29 shows a poor comparison with the experimental data for both monodisperse and polydisperse cases. The trends of the curves are quite different from those of the experimental data.

## 7.2 BISE Experiment

The BISE (Banc de mIse en Suspension par Ecoulement) experiment was performed by Alloul-Marmor (2002) at IRSN/Saclay in France with the purpose of studying the resuspension of non-radioactive polydisperse particles on a surface in fully-developed channel air flow.

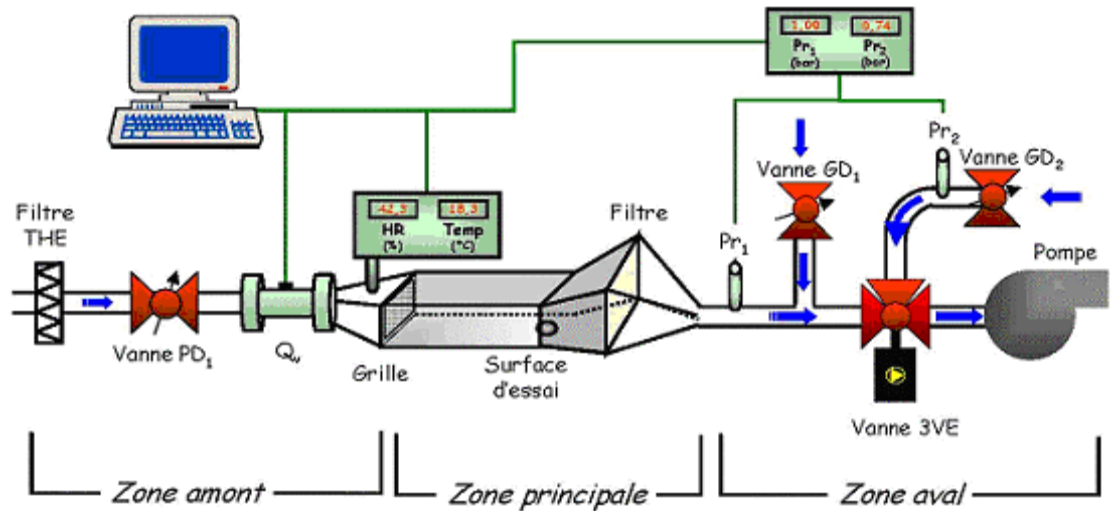

**Figure 30** – Schematic diagram of the BISE experiment apparatus

The principal zone of the installation is composed of two Plexiglas partitions. The one on the left in Figure 30 is a horizontal right-angled parallelepiped conduit 40cm in length, 12cm wide and 7cm high. There was an experimental surface towards the end of the channel, as shown below. The carrier flow was dry air at room temperature with a range of mean velocities from 0.5m/s to 10m/s (friction velocities from 0.04m/s to 0.52m/s). The time interval for each experiment was 900s.

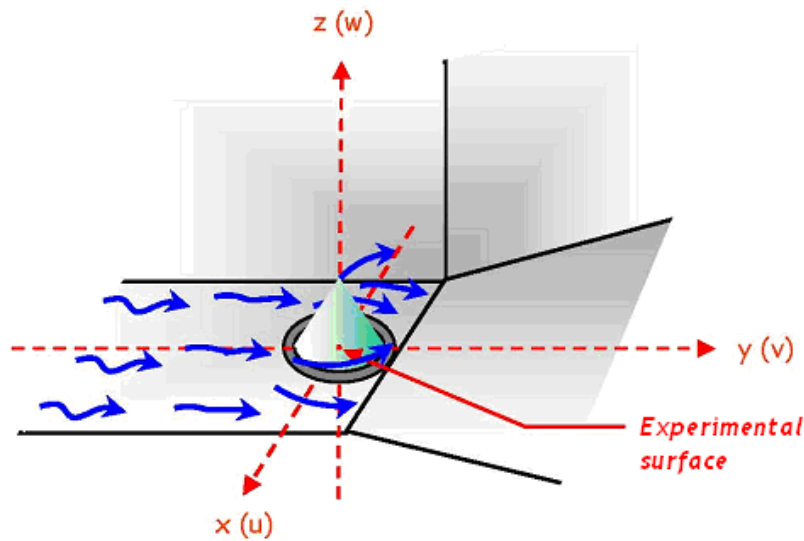

**Figure 31** - Representation of experimental surface

Alumina ($Al_2O_3$) spherical relatively monodisperse particles were used in the experiment. 5 particle sizes were considered from 4.6μm to 58.7μm (MMD) and the resuspension of two of them were compared to R'n'R model predictions. In the model comparison, only one particle size from these two has been chosen (58.7 microns) because the resuspension fraction was very small for the smaller size particles. Also the mass median diameter has been converted to geometric mean diameter, shown in the table below.

| MMD (μm) | GMD (μm) | GSD |
|---|---|---|
| 58.7 | 47.75 | 1.3 |

**Table 6 -** Particle size in the selected BISE experiment

The values of the parameters used in the model calculation are

| Average radius (μm) | Geometric standard deviation of radius | Fluid density ($kg.m^{-3}$) | Fluid kinematic viscosity ($m^2.s^{-1}$) | Surface energy ($J.m^{-2}$) |
|---|---|---|---|---|
| 23.875 | 1.3 | 1.293 | 1.515 x $10^{-5}$ | 0.56 |

**Table 7 -** Parameters in BISE experiment

It is noted that, Biasi's correlation is not appropriate for large size particles. Therefore, the geometric mean of normalized asperity radius and spread factor according to Reeks and Hall (2001) are chosen as 0.01 and 1.5, respectively.

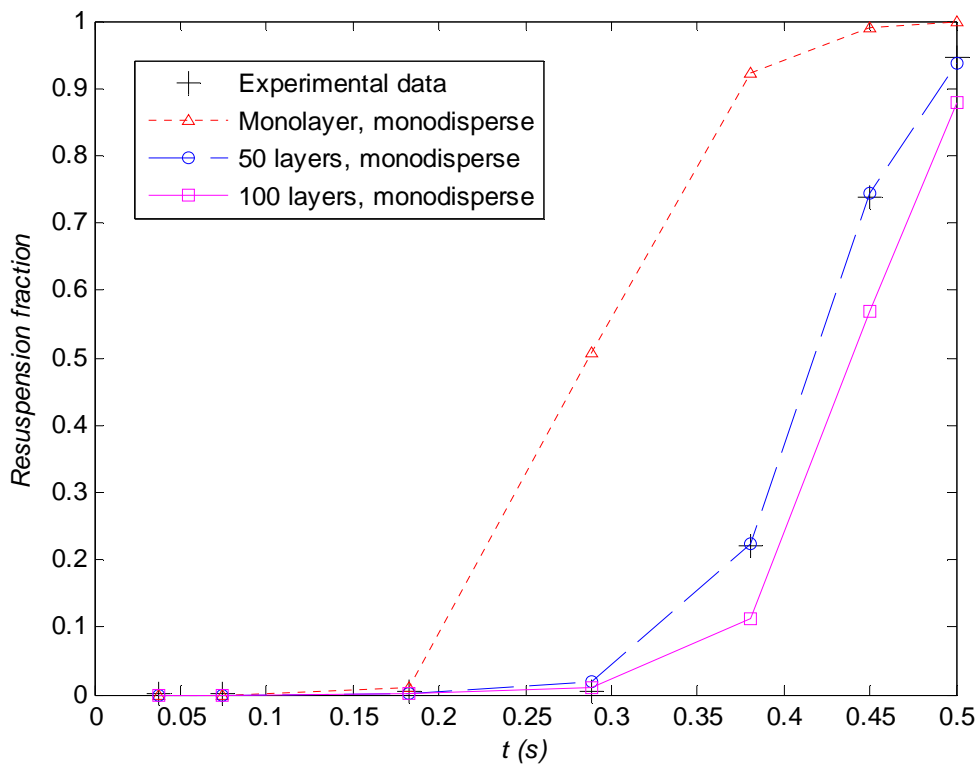

**Figure 32 -** Monodisperse multilayer resuspension predictions vs. BISE experimental results

It is clear from the comparison shown in Figure 32 that the monodisperse multilayer model with 50 layers gives very good results compared to the experimental data. However, when a particle size distribution is included (shown in Figure 33), the polydisperse case gives quite poor comparison with the experimental data.

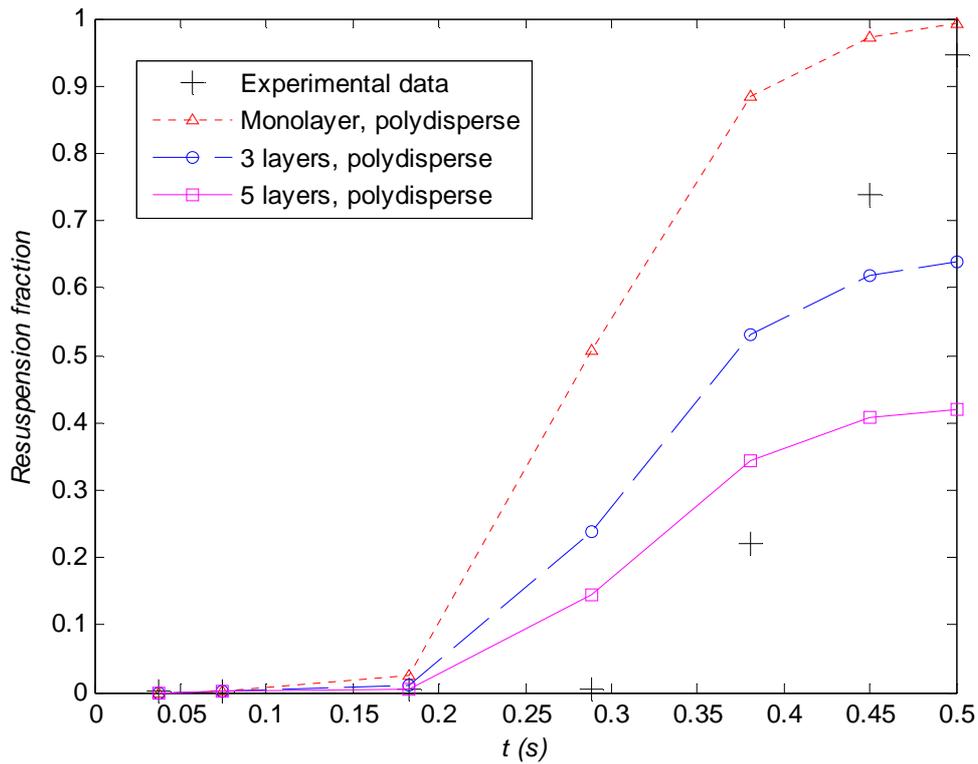

**Figure 33 -** Polydisperse multilayer resuspension predictions vs. BISE experimental results

4.3 Discussion
Several conclusions can be drawn from the validation of the multilayer model.
▪ Both the STORM and the BISE comparisons indicate that multilayer model predictions with small adhesive spread factors give much better comparison with the experimental results. This feature was also concluded by Biasi *et al.* (2001). This may be because within a packed deposit, all the particles have similar surroundings so that the spread in adhesion is reduced when there are a greater number of contacts with other particles (i.e., the deposit is characterized by high average co-ordination number).
▪ In the STORM comparison, the polydisperse case gave a better comparison with the experimental data whereas in the BISE comparison the monodisperse case gave better results. There is a marked difference between the deposit structures in the two experiments: in the STORM test the deposit is formed from a natural deposition phase and would have had a polydisperse character due to the influence of thermophoresis; in the BISE experiment the deposit is a mechanically compacted cone-shaped deposit of relatively monodisperse particles. This may have an influence on the coverage effect of the deposit layers. In other words, due to the fact that in BISE experiments the deposit structure is more compact, the polydisperse resuspension predictions are less applicable to this highly compacted deposit case.
▪ It should be noted that the adhesion parameters (reduction factor and spread factor) basically dominate the multilayer resuspension. Therefore, a better and comprehensive knowledge of these parameters is very important.

## 8. Summary and Conclusions

The multilayer resuspension model that has been developed in this work is a hybrid version of the R'n'R model adapted for application to multilayer deposits based on the Friess and Yadigaroglu (FY) multilayer approach. The deposit is modelled in several overlying layers where the coverage effect (masking) of the deposit layers has been studied; in the first instance a monodisperse deposit with a coverage ratio factor was developed where this was subsequently replaced by the more general case of a polydisperse deposit with a particle size distribution.

The idea behind this new multilayer resuspension model is that the deposited particles are removed by layers; the rate of removal of particles from any given layer depends upon the rate of removal of particles from the layer above which thus acts as a of source of uncovering and exposure of particles to the resuspending flow. This exposure depends upon the surface area exposed to the flow in any given layer and the fraction of that surface area occupied by particles (referred to as the coverage coefficient). It is assumed that when a particle is removed from the layer above only particles with sizes less than that of the particle removed can be resuspended. The influence of coverage and particle size constitute additional features with respect to the original FY model which considered the resuspension of deposit consisting of a regular lattice array of particles of the same size (FY refer to this as a generic model for multilayer resuspension). As in the original FY model a lognormal distribution of adhesive forces between particles is assumed in each layer and that there is no correlation between the adhesive forces between layers. The distribution of adhesive forces and particle-coverage coefficient need not be the same for every layer (indeed, if the layers are numbered sequentially from the top layer then there is good reason to suppose that the adhesive forces increase with increasing layer number). However in this study it is assumed that all 3 parameters are the same in each layer.

We recall that in the case of monolayer resuspension, so long as the exposure time was much greater than timescale of the aerodynamic removal forces the value of this timescale had little influence on the fraction of particles removed. In the case of multilayer resuspension, the value of this timescale is critical to the overall fraction of particles resuspended since it determines the time at which particles in any given layer are exposed to the flow and hence resuspended. In contrast to monolayer resuspension it is the fraction removed as a function of time that is important. In monolayer resuspension most of the particles that are available for resuspension will be removed in less than one second, the remainder being resuspended over much greater times. As an illustration Figure 3 shows the resuspension rate of particles in layers 1, 3, 5, 10 and 20 of a multilayer deposit of particles of the same size. The resuspension rate of the first layer is identical to that of the isolated particle model. For the second and subsequent layers, the initial resuspension rate starts from zero and rises to a maximum during which time most of the particles which are easily removed are resuspended from a given layer (i.e., for these particles the mean effective aerodynamic force > adhesive force), the remainder being removed over a much longer period. The time to reach a maximum may therefore be regarded as a delay time for the particles in a given layer to be exposed by removing particles from all the layers above it. This behaviour results in a dramatic difference in the fraction resuspended compared with that for a monolayer for the same exposure time. Figure 6 shows for instance the resuspension fraction of a multilayer deposit consisting of 100 layers. It shows that after 100s, nearly 80% of the monolayer deposit is removed compared to around 3% of the 100 layers deposit.

The spread factor of the adhesive force distribution also has a great influence on multilayer resuspension. As a typical example, Figure 10 shows the half-life (time to resuspend 50% particles) for different spread factors. As the spread factor increased, the half-life increases dramatically as the deposit becomes thicker. We also compared the resuspension prediction using the non-Gaussian model for the resuspension rate constant (based on the DNS results) with those based on a Gaussian model (with the same values of $\omega, f_{rms}, \langle F \rangle$). As with the case of mono-layer resuspension, Figure 13 shows that the non-Gaussian model always results in more resuspension than the Gaussian case but, in contrast, as time increases the difference becomes greater and greater. This difference reflects the difference in the normalized distributions for the aerodynamic drag force where the non-Gaussian distribution (used throughout this study) is significantly greater in the wings than that of the Gaussian distribution. It was noted that when both models gave the same value for the resuspension fraction (around $10^{-3}$s in the plot) this value was independent of the layer thickness.

In studying the resuspension of a multilayer deposit composed of polydispersed particles the model predictions for the resuspension fraction were compared with those of a multilayer deposit composed of monodisperse particles with the same geometric mean size. As an example, Figure 21 shows the ratio of the resuspension fraction of the monodisperse particles with that of the polydisperse particles as a function of layer thickness ($L = 1 \sim 100$) (reduction and spread factors are calculated by Biasi's correlation). All the curves start with similar values due to the dominance of resuspension of the top layers; also in the short term ($<10^{-4}$s) the resuspension of the polydisperse particles is greater than that of the monodisperse particles since the 50% of the polydisperse particles with sizes > that of the monodisperse particles are easier to resuspend. As the number of layers increases the difference between the monodisperse and polydisperse particles becomes significant. For 10 layers, the resuspension fraction of the monodisperse particles after 100s is ~4.5 times that of polydisperse particles, and for the 100-layers case, the ratio increases to around 5 after 100s.

As a final conclusion, it has been seen that the present work significantly improves prediction of the kinetics of the resuspension of fine particles from multilayer deposits. In particular, with *a priori* information regarding the resuspension, it can be used to draw inferences concerning the structure of such deposit. Therefore, to improve its predictive capability, an ultimate improvement would encompass modelling of the structuring of such deposits as they form including clustering: this information fed into the present model would constitute a complete predictive model of the resuspension phenomenon.

**Acknowledgement**
This work is part of the lead author's PhD research initiated and funded by IRSN and undertaken in collaboration with the University of Newcastle (UK) and the Ecole Centrale de Lyon (F).